\documentclass[aps,prl,twocolumn,showpacs,epsfig,floats]{revtex4-1}
\usepackage{amsmath}
\usepackage{color}
\usepackage{graphicx}
\usepackage{epsfig}
\usepackage{amsfonts}
\usepackage{mathrsfs}
\usepackage{mathtools}
\usepackage{amsmath}
\usepackage{float}
\usepackage{physics}
\usepackage{hyperref}
\hypersetup{
  colorlinks=true,
  citecolor=blue,
  linkcolor=blue,
  urlcolor=blue}

\begin{document}

\title{Self-trapping phenomenon, multistability and chaos in open anisotropic Dicke dimer
}

\author{G. Vivek, Debabrata Mondal, Subhadeep Chakraborty, and S. Sinha}
\affiliation{Indian Institute of Science Education and
	Research-Kolkata, Mohanpur, Nadia-741246, India}
	
\begin{abstract}
	We investigate semiclassical dynamics of a coupled atom-photon interacting system described by a dimer of anisotropic Dicke model in the presence of photon loss, exhibiting a rich variety of non-linear dynamics. Based on symmetries and dynamical classification, we characterize and chart out various dynamical phases in a phase diagram.
	A key feature of this system is the multistability of different dynamical states, particularly the coexistence of various superradiant phases as well as limit cycles.
	Remarkably, this dimer system manifests self-trapping phenomena, resulting in a photon population imbalance between the cavities. Such a self-trapped state arises from a saddle-node bifurcation, which can be understood from an equivalent Landau-Ginzburg description. Additionally, we identify a unique class of oscillatory dynamics ``{\it self-trapped limit cycle}", hosting self-trapping of photons. The absence of stable dynamical phases leads to the onset of chaos, which is diagnosed using the saturation value of the decorrelator dynamics. 
	Moreover, the self-trapped states can coexist with chaotic attractor, which may have intriguing consequences in quantum dynamics. Finally, we discuss the experimental relevance of our findings, which can be tested in cavity and circuit quantum electrodynamics setups.
	
\end{abstract}

\date{\today}
\maketitle
{\it Introduction:}
Nonequilibrium dynamics of the quantum many body systems has become a subject of intense research in recent years \cite{rev_noneq_1,rev_noneq_2,Moore_noneq_3,Huse_noneq_4}. 
The advancement of ultracold atomic system as well as cavity and circuit  quantum electrodynamics (QED) paves the way for exploring a range of nonequilibrium phenomena, including various types of phase transitions and thermalization processes \cite{rev1,rev2,rev4,Serge_Haroche,Carusotto_Review,Helmut_Ritsch,Esslinger_rev,Zollar_rev,Steven_Girvin_1,Houck, Review_ergodicity_1,Blatter,Le_Hur,knap1,jcref7,fazio1}. These atom-photon interacting systems inherently exhibit dissipative effects due to various loss processes, providing a platform to study the rich dynamical behavior of open quantum systems \cite{Daley_dissipation,Mueller_dissipation,Hendric,Keeling_Dicke,Hemmarich,Dis_tran4,Dis_tran5,Dis_tran6, Dis_tran7,Dicke_5,Dicke_6,Dis_tran9, Dis_tran10,Dis_tran11}.
Understanding nonequilibrium phases of open quantum systems and nature of the transition between them has attracted significant interest \cite{Dis_tran1,Dis_tran8,Chitra,Cuiti_dist_trans_1,Zollar_dist_trans,dissipative_transition_CQED,Cuiti_dist_trans_2, Milies_dist_trans}. 
Seminal experiments have demonstrated the formation of exotic phases and nonequilibrium transitions by coupling cold atoms with cavity field \cite{supersolid_1,supersolid_2, Hemmarich, A_M_Rey_cavity_nature,A_M_Rey_cavity_exp, Dissipative_Rydberg,B_L_Lev,ADM_1,Dicke_exp_1,Dicke_exp_2,Dicke_exp_3,Stojan_Rebic}, as well as in the circuit QED array \cite{dissipative_transition_CQED}. 
Additionally, oscillatory dynamics in the form of quantum limit cycles (LC) typically occur in the presence of dissipation, which leads to the intriguing time crystalline phase \cite{LC_Disssipative_Rydberg,Limit_cycle_Hemmerich_2,continous_time_crystal,Quantum_vanderpol,quantum_LC_4,quantum_LC_5,quantum_LC_6, Dicke_correlator}.
On the other hand, the signature of chaos and its manifestation in the context of ergodicity and thermalization is a well studied area for isolated quantum system \cite{rev_noneq_2,Review_ergodicity_1}, which is however less explored for open quantum systems in the presence of dissipation \cite{Haake_chaos_1,Casati_chaos_1,Zyczkowsky,Shepelyansky,Prosen_3,BHT_chaos,Deb_TCH,Minganti,Russomanno}.
Furthermore, the collective nature of these atom-photon systems allows to analyze the dynamics semiclassically, offering a pathway to probe the onset of chaos in open quantum systems \cite{Casati_chaos_1,QPT_Chaos_Brandes,Sudip_review,Graham_EPL,Kolovsky}.

The light-matter interacting system in a cavity containing multiple two level atoms can  effectively be described by the anisotropic Dicke model \cite{ADM_1,ADM_2,Dis_tran4}. The paradigmatic Dicke model \cite{QPT_0} and its variants exhibit a plethora of fascinating phenomena, which include various types of phase transitions, chaos and thermalization \cite{ QPT_Chaos_Brandes,Haake_2012,Lea_2023,A_M_Rey_Dicke_1,Scar_Dicke,Scar_sudip, Dicke_1,Dicke_2}.
In the present work, we consider nonequilibrium dynamics of two Josephson coupled cavities containing many atoms described by a dimer of anisotropic Dicke model. The photon hopping between the cavities can be engineered \cite{JCH_Plenio,npj_hopping}, as already demonstrated in the circuit QED experiments.
In the presence of photon loss, this open anisotropic Dicke dimer (ADD) model exhibits a rich variety of dynamical phenomena as well as the formation of different nonequilibrium phases and transitions between them. 
In this letter, we focus on a new class of steady-state and oscillatory dynamics that lead to self-trapping phenomena of light even in the presence of photon loss, detectable from the photon number imbalance between the cavities.
%
Self-trapping phenomena have attracted interest since their observation in cold atom systems \cite{Obarthalar_1,Oberthalar_2,AC_DC}, and have been explored both theoretically and experimentally in other systems as well \cite{Self_trapping1,JC_dimer_expt,Shenoy1,Shenoy2,Sebastian_ST, JCD_Houck, Vivek_JCD, Manas_TCD}. However, while such states typically have a finite lifetime in an open environment \cite{Self_trapping1,Shenoy_damping,Scar_sudip}, the present model (ADD) displays a stable self-trapping phenomena even in the presence of photon loss. 
%
Furthermore, we investigate the dissipative chaos in ADD and its physical consequences.

\begin{figure}
	\includegraphics[width=\columnwidth]{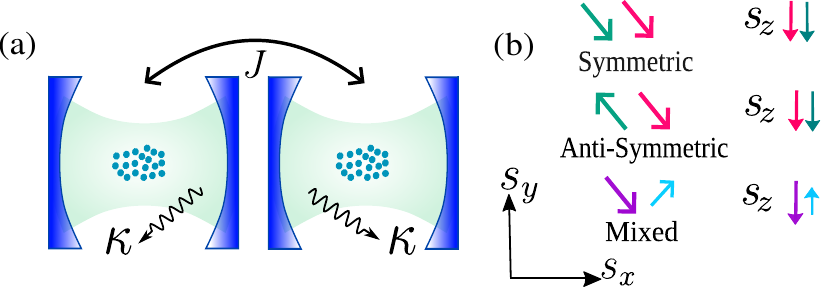}
	\caption{ {(a) Schematic of the open anisotropic Dicke dimer. (b) Spin orientation representing different dynamical classes.}}
	\label{fig1}
\end{figure}

{\it Model and semiclassical analysis:}
We consider two coupled cavities each containing $N$ two level atoms interacting with a single cavity modes (as shown in Fig.\ref{fig1}(a)) which can be described by the Dicke dimer model with the following Hamiltonian,
\begin{small}
	\begin{eqnarray}
		\hat{\mathcal{H}} \!\!&=&\!\!\! -J \left( \hat{a}_{\mathrm L}^\dagger \hat{a}_{\mathrm R} + \hat{a}_{\mathrm R}^\dagger \hat{a}_{\mathrm L}\right)\!+\!\!\sum_{i=\rm L, R} \left[\omega \hat{a}_i^\dagger\hat{a}_i + \omega_0 \hat{S}_{zi}\right. \nonumber \\
		&&\!\!\!+ \left.\frac{\lambda_-}{\sqrt{2S}}\left(\hat{a}_i^\dagger \hat{S}_{i-} + \hat{a}_i \hat{S}_{i+}\right) 
		+\frac{\lambda_+}{\sqrt{2S}}\left(\hat{a}_i \hat{S}_{i-} \!+\! \hat{a}_i^\dagger  \hat{S}_{i+}\right) \!\right]
		\label{ADM}
	\end{eqnarray}
\end{small}
where the site index $i=\rm L(R)$ represents left(right) cavity, $\hat{a}_{i}$ annihilates photon mode with frequency $\omega$, and 
$J$ is the hopping amplitude of the photons between the cavities.
Each cavity contains $N$ two level atoms with energy gap $\omega_0$, that are collectively represented by large spins $\hat{\vec{S_{i}}}$ with magnitude $S =N/2$, and $\lambda_{\pm}$ are atom-photon coupling strengths of the anisotropic Dicke model (ADM), which has been extensively studied both classically and quantum mechanically \cite{Dis_tran4,Dis_tran5,Dis_tran6,Dis_tran7}.

%
%
For this system with $S\gg 1$, the scaled operators $(\hat{x}_{i} + \imath \hat{p}_{i})/\sqrt{2} = \hat{a}_{i}/\sqrt{S}$ and $\hat{\vec{s}}_{i} = \hat{\vec{S}}_{i}/S$ become classical, as they satisfy, $[\hat{x}_{i},\hat{p}_{i}]= \imath/S$ and $[\hat{s}_{ai}, \hat{s}_{b i}] = \imath \epsilon_{abc} \hat{s}_{c i}/S$, where $1/S$ plays the role of reduced Planck constant.

Realistically, photon loss from the cavities is inevitable, resulting in non-unitary time evolution of the density matrix (DM) of this open ADD model, which can be described by the Lindblad master equation \cite{Breuer},
\begin{equation}
\frac{d\hat{\rho}}{dt} = - \iota [\hat{H},\hat{\rho}] + \sum_{i= \rm L, R } \left(2\hat{L}_{i}\hat{\rho} \hat{L}_{i}^\dagger - \{\hat{L}_{i}^\dagger \hat{L}_{i}, \hat{\rho}\}\right),
\label{ME}
\end{equation}
where $\hat{L}_{i} = \sqrt{\kappa} \hat{a}_{i}$ account for the most dominant source of dissipation in cavity QED setups due to the photon decay with amplitude $\kappa$ \cite{ADM_1,Dicke_exp_1,Dicke_exp_2,Dicke_exp_3, Limit_cycle_Hemmerich_2}. 
%
In the absence of the coupling $J$, the above master equation has already been used to describe the single cavity experiments \cite{ADM_1,Dicke_exp_1,Dicke_exp_2,Dicke_exp_3, Limit_cycle_Hemmerich_2}.
Apart from relaxation dynamics, the nonlinearity in the Lindblad master equation can also drive the open quantum system to a variety of fascinating stationary and non-stationary states \cite{Drossel}.
Throughout the paper, we set $\hbar,k_B=1$ and scale the coupling strengths (time) by $J$ $(1/J)$. 
The time evolution of the average value of any operator $\hat{O}$ can be obtained using the relation $\frac{d\langle \hat{O} \rangle}{dt} = 
{\rm Tr}(\hat{O} \dot{\hat{\rho}})$. Within the mean-field approximation, the expectation of the product of operators can be decomposed as $\langle \hat{A} \hat{B}\rangle = \langle \hat{A} \rangle \langle \hat{B} \rangle$, which is valid for $ S \rightarrow \infty$ \cite{Sudip_review,Dis_tran8,Dis_tran11, Dicke_correlator}. 
For the typical experimental setups with $N\approx 10^4$, the Eq.\eqref{ME} can be analyzed semiclassically \cite{ADM_1,Dicke_exp_1,Dicke_exp_2,Dicke_exp_3,Limit_cycle_Hemmerich_2} and the dynamics of the scaled classical observables is described by following equations of motion (EOM),
%
\begin{subequations}
	\begin{eqnarray}
         \dot{\alpha_i} &=& -(\kappa+\imath\omega) \alpha_i-\frac{\imath}{\scalebox{0.8}{$\sqrt{2}$}}(\lambda_-s_{i}^-+\lambda_+s_{i}^+)+\imath J\alpha_{\bar{i}} \\
         \dot{s}_{i}^+ &=& \imath\omega_0 s_{i}^+-\imath\scalebox{0.8}{$\sqrt{2}$} s_{zi}(\lambda_-\alpha_i^*+\lambda_+\alpha_i)\\
         \dot{s}_{zi} &=& -\frac{\imath}{\scalebox{0.8}{$\sqrt{2}$}}[ \lambda_-(\alpha_i s_{i}^+-\alpha_i^*s_{i}^-)+\lambda_+(\alpha^*_i s_{i}^+-\alpha_is_{i}^-)]\quad
	\end{eqnarray}
\label{EOM}
\end{subequations}
where $\bar{i}\ne i$ and $\alpha_i = (x_i+\imath p_i)/\sqrt{2}$ with scaled photon number $n_i=|\alpha_i|^2$. For convenience, we define the dynamical variables by an array $\mathcal{X}_i=\{Q_{i},s_{zi}\}$, with $Q_{i}=\{x_{i},p_{i},s_{xi},s_{yi}\}$. 
The above EOM obey 
$\mathbb{Z}_{2}$ parity symmetry corresponding to $\{Q_{\rm L}, Q_{\rm R}\} \rightarrow \{-Q_{\rm L}, -Q_{\rm R}\}$ and $s_{z i}\rightarrow s_{zi}$. 
Additionally, the EOM remain invariant when the site indices are interchanged, $\rm L \leftrightarrow \rm R$, exhibiting exchange symmetry. Both symmetries are relevant for characterizing different dynamical states.
 
To understand the overall dynamics and different nonequilibrium phases, we analyze the steady states (fixed points) $\mathcal{X}^*_i$ from the EOM and their stability, as  discussed in the supplementary material \cite{SM}. 
Introducing the dynamical variables, $\mathcal{X}_{\pm}=\{\frac{Q_{\rm L}\pm Q_{\rm R}}{2},\frac{s_{z{\rm L}}\pm s_{z{\rm R}}}{2}\}$ helps to classify the phase space dynamics in the following categories: \\
{\it \textbf{i)  Symmetric}}: as evident from the EOM (see Eq.\eqref{EOM}), $\{Q_-,s_{z-}\}=0$ remains as a fixed point, which can, in turn, impose a constraint and the dynamics is governed by the variables $\mathcal{X}_+$ only, within a restricted phase space.\\
{\it \textbf{ii) Anti-symmetric}}: on the other hand, the fixed point $\{Q_+,s_{z-}\}=0$ confines the dynamics to a restricted subspace of the dynamical variables $\mathcal{X}_-$.\\
Note that $s_{z{\rm L}}=s_{z{\rm R}}$ for both classes, however, the projection of spins in the $x-y$ plane remains parallel (anti-parallel) for symmetric (anti-symmetric) class, as depicted in Fig.\ref{fig1}(b) \cite{foontnote}.\\
{\it \textbf{iii)  Mixed}}: generally, the dynamics may not be restricted to the symmetric or anti-symmetric classes only, especially when the corresponding fixed points become unstable, resulting in their mixing.


\begin{figure}[t]
	\includegraphics[width=\columnwidth]{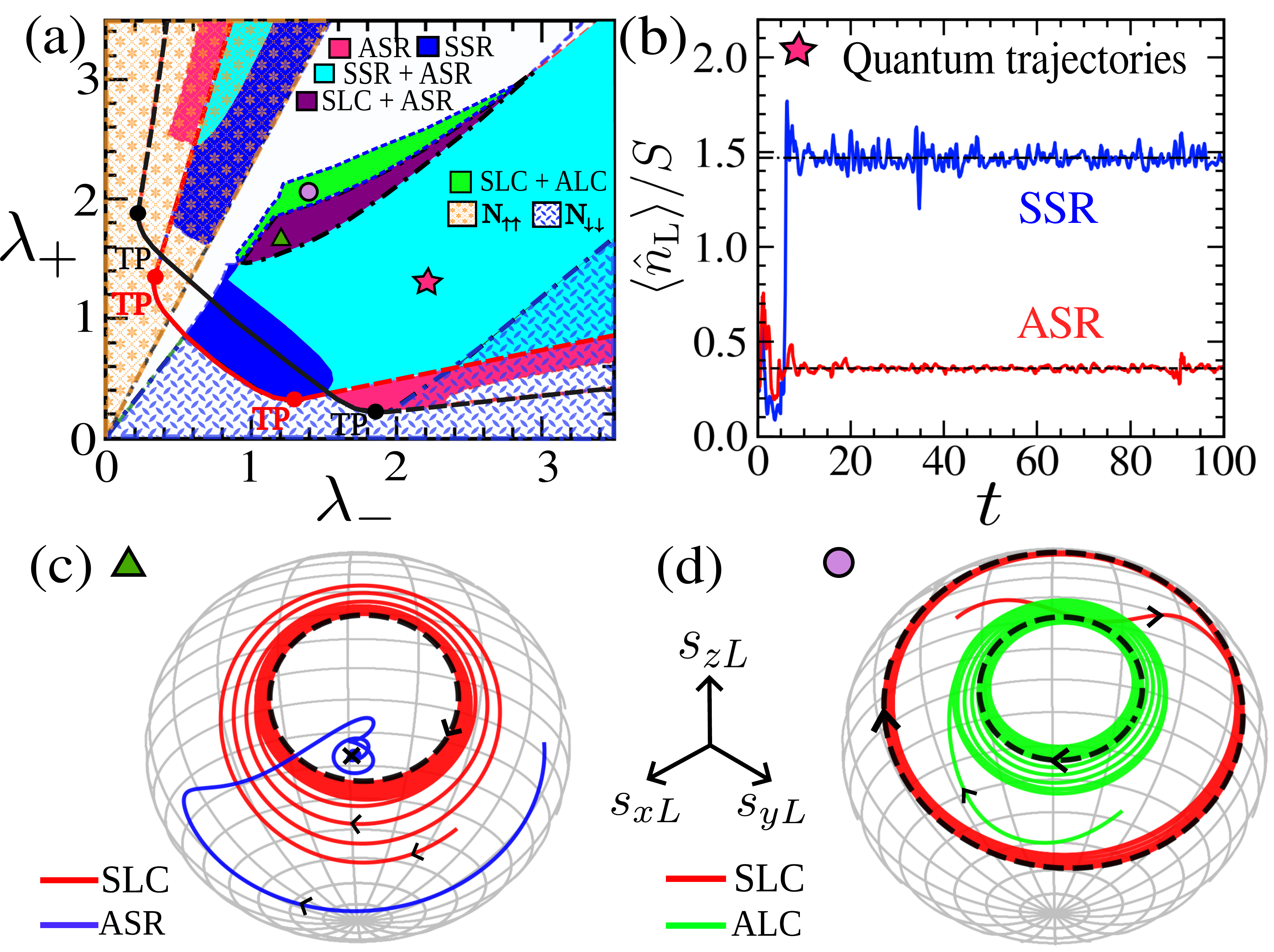}
	\caption{ {\it Symmetric and anti-symmetric superradiant phases and limit cycles:} (a) Classical phase diagram of stability regions of SSR (blue), ASR (pink) phases, their bistability (cyan) and the coexistence of ASR, SLC (violet) and SLC, ALC (lime), as well as  
	the NP$_{\uparrow \uparrow}$ (yellow stars) and NP$_{\downarrow \downarrow}$ (blue diamonds). Transition of SSR (ASR) from NP$_{\downarrow \downarrow}$ represented by pink (black) lines.
	Normal phases NP$_{\uparrow \downarrow}$, NP$_{\downarrow \uparrow}$ are always unstable.
	%
	%
	(b) {\it Quantum dynamics within stochastic wavefunction approach for $S=5$:} average scaled photon number $\langle \hat{n}_{\rm L}\rangle/S$ for different quantum trajectories converge to the classical steady state of either SSR or ASR in the bistable regime at $\lambda_-=2.2,\lambda_+=1.2$ (magenta star in (a)).
	The classical spin dynamics depicting the coexistence of (c) ASR and SLC at $\lambda_-=1.2,\lambda_+=1.64$ (green triangle) and (d) both limit cycles SLC, ALC for $\lambda_-=1.4,\lambda_+=2.0$ (violet circle).
	All energies (time) are measured by $J (1/J)$. We set $\hbar,k_B = 1$ and $\omega=2.7,\omega_0=0.9,\kappa = 0.9$ for all figures.}
	\label{fig2}
\end{figure}

Next, we investigate the steady states characterized by the above mentioned dynamical classes and their stability. Interestingly, the EOM corresponding to the symmetric and anti-symmetric classes reduce to that of a single ADM with effective photon frequency $\omega_{\mp}=\omega\mp J$ respectively. However, their stability can be quite different as the fluctuations are not restricted to any particular symmetry class.
First, we identify two stable normal phases ({\rm NP$_{\downarrow\downarrow}$},{\rm NP$_{\uparrow\uparrow}$}) with zero photon number and the spin polarization $s^*_{z+}=\mp1$ respectively. 
By changing the coupling strength, the normal phase NP$_{\downarrow\downarrow}$ undergoes a transition to parity symmetry broken superradiant phase with non-zero photon number. In contrast to the single ADM, in this case, both the symmetric (SSR) and anti-symmetric (ASR) superradiant phases emerge from NP$_{\downarrow\downarrow}$, each belonging to their respective dynamical class. 
The stability regimes of the normal phases and the formation of stable SR phases as well as their transitions in the parameter space are summarized in the phase diagram, given in Fig.\ref{fig2}(a).
The nature of the SR transition changes from a pitchfork to a saddle-node bifurcation depending on the parameter regime, as depicted by solid and dashed lines in Fig.\ref{fig2}(a) respectively.
%
%
This scenario of nonequilibrium transition can be captured analytically from an effective Landau-Ginzburg potential $\mathcal{F}(m)=\frac{a}{2}m^2+\frac{b}{4}m^4+\frac{c}{6}m^6$, with the order parameter $m^2=1+s_{z+}$  \cite{SM}. For $b>0$, the continuous transition (pitchfork) changes to a first-order transition (saddle-node)  at the tricritical point (TP) with $a,b=0$.

Interestingly, both the SSR and ASR phases can coexist in a large region of the phase diagram, exhibiting bistability (Fig.\ref{fig2}(a)).
However, their basins of attraction do not overlap, 
guiding the appropriate choice of initial conditions for the  detection of the respective phases. 

At the upper boundary of the bistable regime (dashed-dotted line) in the phase diagram in Fig.\ref{fig2}(a)), the SSR phase becomes unstable and undergoes a Hopf bifurcation, giving rise to a pair of limit cycles belonging to the symmetric dynamical class (SLC), around two unstable branches of SSR phase. Further increasing $\lambda_+$, the ASR phase also becomes unstable and undergoes a similar Hopf bifurcation, giving rise to another pair of anti-symmetric LC (ALC). 
Interestingly, the SLC can coexist with either the ASR phase or the ALC in different parameter regions, giving rise to fascinating bistability. In these regions, trajectories converge to different attractors depending on initial conditions, as shown by the spin dynamics on the Bloch sphere (Fig.\ref{fig2}(c,d)).
%
Eventually, these limit cycles become unstable, leading to a quasi-periodic motion and chaos.
\begin{figure}[b]
	\includegraphics[width=\columnwidth]{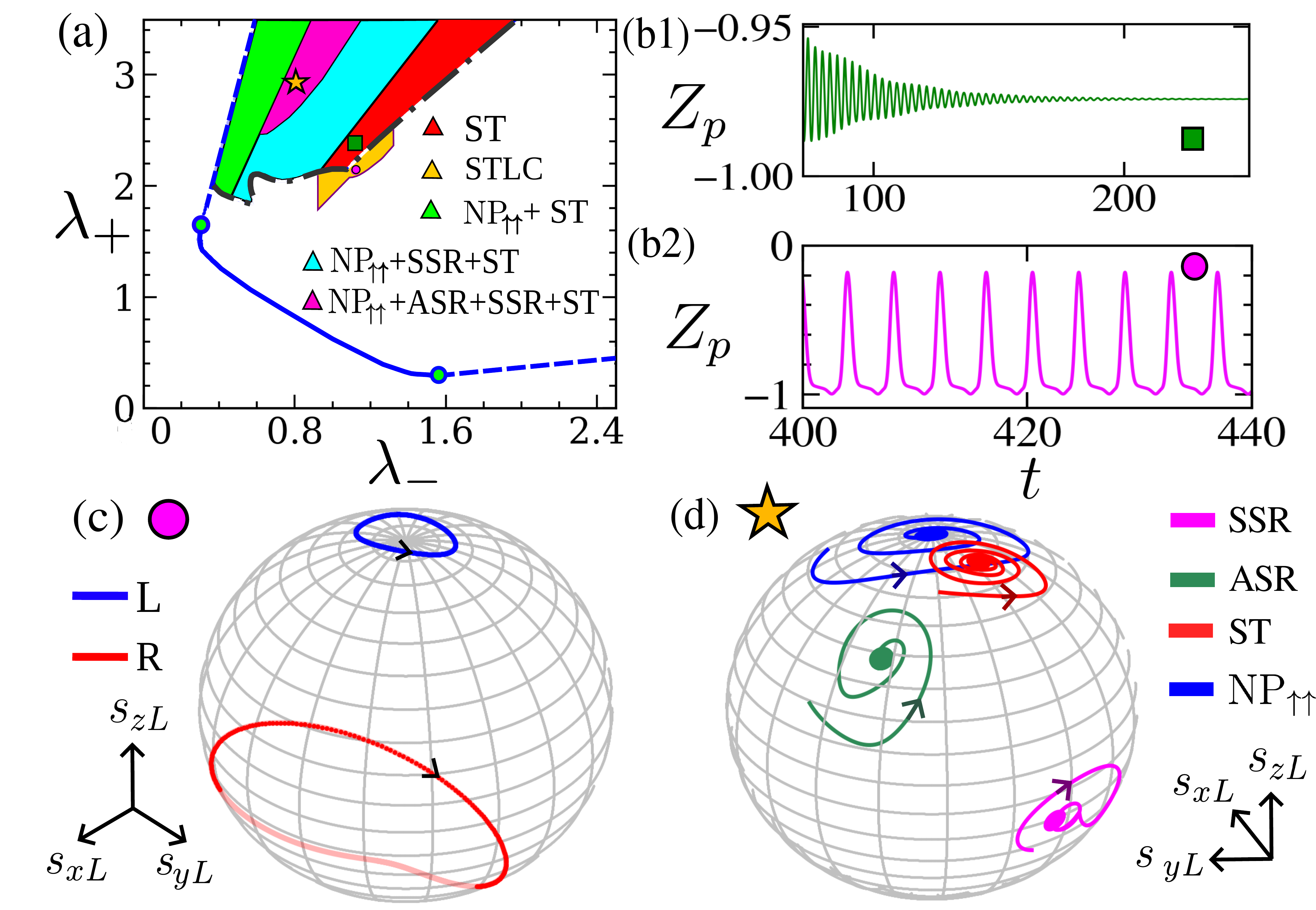}
	\caption{{\it Self-trapping phenomenon:} (a) Stability regions of steady states (see the text for details). 
	%
	The solid and dashed blue lines indicate the critical lines for the  ST state.
	The stable ST state is bounded by the Dashed dotted black line. 
	Dynamics of the photon imbalance $Z_p$ of (b1) ST state at $\lambda_-=1.15$, $\lambda_+=2.4$ (green square in (a)) and (b2) the STLC for $\lambda_-=1.15$, $\lambda_+=2.12$ (magenta circle in (a)). (c) Shape of the STLC (on the Bloch sphere) in the left (blue) and right (red) cavities at the parameter same as panel (b2). 
	(d) Spin Dynamics reaching different steady states based on initial conditions in the multistability regime at $\lambda_-=0.8$, $\lambda_+=2.9$ (orange star in (a)).
}
	\label{fig3}
\end{figure}

{\it Self-trapping phenomenon:} To this end, we focus on another type of mixed class of dynamics, which breaks the exchange symmetry between two cavities, leading to a fascinating self-trapping phenomenon of photons. 
In the self-trapped (ST) state, the photon population in one of the cavities is larger, resulting in a non-vanishing photon number imbalance $Z_p = \frac{n_{\rm L}-n_{\rm R}}{n_{\rm L}+n_{\rm R}}$ and unequal spin polarization $|s^*_{z{\rm L}}|\ne |s^*_{z{\rm R}}|$ \cite{SM}. The signature of the stable ST state can be detected from the dynamics of the population imbalance $Z_p$, which saturates to a non-vanishing value (see Fig.\ref{fig3}(b1)).
The nature of the transition leading to the ST state and its regime of stability are summarized in Fig.\ref{fig3}(a). To investigate the formation of the ST steady state, we find that it stems from either of the unstable normal phases NP$_{\uparrow\downarrow}$ or NP$_{\downarrow\uparrow}$ with $|s^*_{z-}|=1$ and $s^*_{z+}=0$.
%
Similar to the superradiant transition, the ST state originates from the aforementioned normal phases through a pitchfork bifurcation (continuous transition) along a critical line (solid blue line in Fig.\ref{fig3}(a)),  which eventually changes its nature to a saddle-node bifurcation (dashed blue line in Fig.\ref{fig3}(a)), above a tricritical point. 
The phase boundary and the nature of transitions to the ST state can also be understood by constructing an effective LG potential
$\mathcal{F}(m)$, where $m^2=1-s_{z-}$ serves as an order parameter (see \cite{SM} for details). 
We emphasize that the stable ST phase originates only from the saddle-node bifurcation and exists in a specific region of the phase diagram, as shown in Fig.\ref{fig3}(a).
%
Another type of ST state originating from the unstable SR phase, always remains unstable, which is omitted from our discussion.

Interestingly, within a large region of the phase diagram shown in Fig.\ref{fig3}(a), the stable ST state coexists with both the stable superradiant phases SSR and ASR, as well as with the stable normal phase NP$_{\uparrow\uparrow}$, leading to an intriguing multistability of nonequilibrium phases. For such coexisting fixed points, the different phases can be probed dynamically depending on the initial conditions, as shown in Fig.\ref{fig3}(d). The different fixed points in this multistable region have a non-overlapping basin of attraction \cite{SM}, which guides the choice of the initial state suitably to observe the self-trapping phenomenon.
Notably, the stable self-trapped state persists for a wide range of decay rates $\kappa$ \cite{SM}, ensuring the robustness of the phenomenon even in the presence of dissipation.

Apart from the ST fixed point, we also identify a unique oscillatory dynamics leading to the self-trapping phenomenon, which we dub as {\it self-trapped limit cycle} (STLC).
Such a limit cycle exists within a narrow region close to the boundary of the ST state (see Fig.\ref{fig3}(a)), which is typically observed by initializing the dynamics close to the unstable normal phase NP$_{\uparrow\downarrow}$. 
As seen from Fig.\ref{fig3}(c), the spin dynamics corresponding to the self-trapped limit cycle leads to the formation of two distinctly different contours on the Bloch sphere of two cavities, exhibiting the exchange symmetry breaking.
%
Furthermore, the self-trapping phenomenon corresponding to STLC is evident from the dynamics of photon imbalance $Z_p$, which shows periodic oscillation with non-vanishing average value (see Fig.\ref{fig3}(b2)).
\begin{figure}[b]
	\includegraphics[width=\columnwidth]{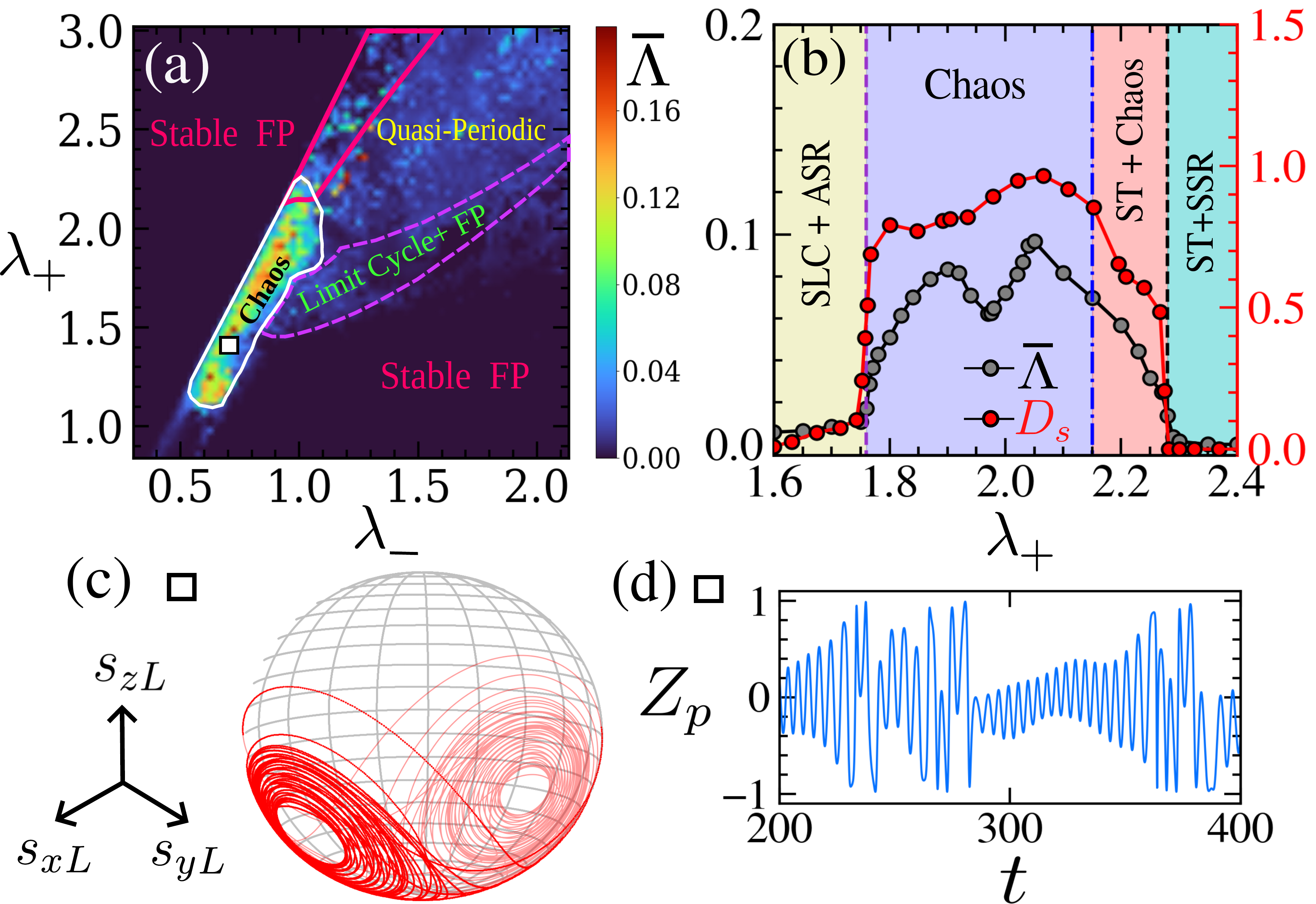}
	\caption{ {\it Dissipative Chaos:} (a) Average Lyapunov exponent $\bar{\Lambda}$ on the phase diagram as a color scale. (b) Average Lyapunov exponent (left axis) and the saturation value of decorrelator $D_s$ (right axis) as a function of $\lambda_+$ for $\lambda_-=1.0$, exhibiting different dynamical regimes. Chaotic dynamics of (c) the spin in one cavity over Bloch sphere and (d) the photon imbalance $Z_p$ at the interaction strengths $\lambda_-=0.7,\lambda_+=1.4$. 
	}
	\label{fig4}
\end{figure}

{\it Chaotic dynamics :} In addition to the various steady states and limit cycles, the ADD model also exhibits dissipative chaos in a narrow region of the phase diagram (see Fig\ref{fig4}(a)). Due to the instability of the fixed points, the region of regular dynamics shrinks significantly, which leads to the onset of chaos.
The chaos stems from the instability of the symmetric limit cycles (SLC), leading to the  chaotic motion of spins revolving around two pairs of unstable SSR fixed points (see Fig.\ref{fig4}(c)), resembling the behavior of the Lorentz system \cite{Lorentz,Strogatz}.

We identify the chaotic region (depicted in Fig.\ref{fig4}(a)) from the positive average Lyapunov exponent (LE) $\bar{\Lambda}$, computed for an ensemble of phase space points \cite{Lichtenberg,Lyapunov}. 
%
Furthermore, we study the decorrelator \cite{Decorr1,Decorr2,Decorr3,Decorr4,Decorr5,Decorr6,Decorr7} dynamics of spin $D(i,t)=1-\langle {\bf s}_i^a\cdot {\bf s}_i^b\rangle$, where the initial configurations $a,b$ differ by a small perturbation, $\mathbf{s}^{b} = \mathbf{s}^{a}+\delta \mathbf{s}$, and the averaging is performed over the initial states. 
%
In the chaotic regime, the decorrelator initially grows and finally saturates below unity, whereas it remains vanishingly small for regular phase space dynamics. As evident from Fig.\ref{fig4}(b) the saturation value of the decorrelator $D_s$ serves as an alternate measure to quantify overall chaoticity, which can efficiently identify the chaotic region in the parameter space. 
%
Fig.\ref{fig4}(a) depicts the gradual transition of the chaotic region into quasi-periodic motion, characterized by a peaked distribution of frequencies and a small value of the Lyapunov Exponent (LE). 

\begin{figure}[t]
	\includegraphics[width=\columnwidth]{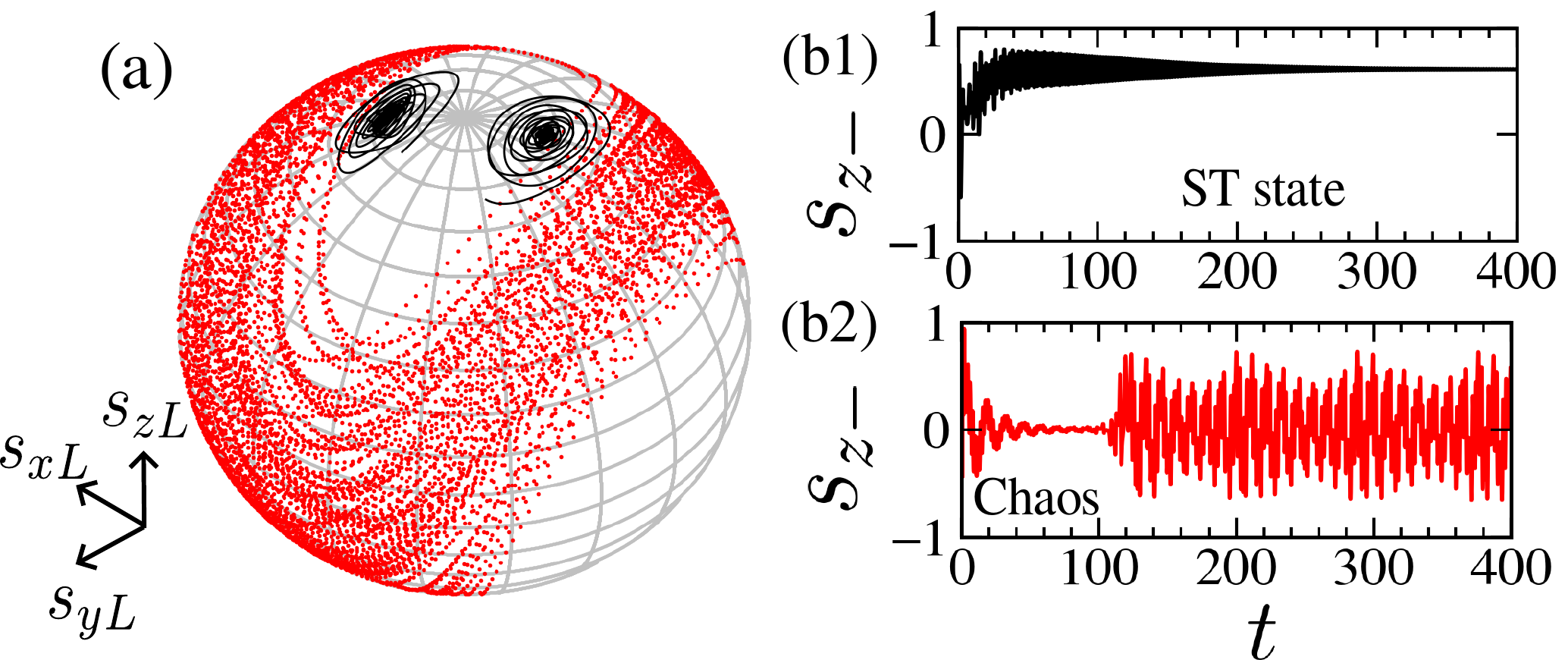}
	\caption{ {\it Coexistence of chaos and self-trapped state:} (a) Trajectories on the Bloch sphere converging to the two branches of ST state (black) corresponding to the parity symmetry and chaotic attractor (red). Dynamics of $s_{z-}$ for (b1) of ST state and (b2) chaotic regime. We set $\lambda_-=1.0$ and $\lambda_+=2.25$ from the region marked in Fig.\ref{fig4}(b).}
	\label{fig5}
\end{figure}
In the current scenario of dissipative chaos, the attractor occupies a finite region of phase space.
In consequence, the ST steady states can coexist with chaos, as shown over the Bloch sphere in Fig.\ref{fig5}(a). This remarkable coexistence phenomena can be probed dynamically, as the trajectory initialized with zero photon number and $s_{z{\rm L}}=1,s_{x{\rm R}}=\pm1$ converges to the ST state, while chaotic dynamics can be observed for initial $N_{\downarrow \downarrow}$ state (compared in Fig.\ref{fig5}(b1,b2)). 
Note that the degree of chaos and self-trapping phenomenon can also depend on the detuning $\delta=\omega-\omega_0$  \cite{SM}. 

{\it Quantum dynamics}:  To this end, we study the semiclassical dynamics using truncated Wigner approximation (TWA) considering a large number of atoms, where a stochastic noise is added due to photon loss  \cite{continous_time_crystal,Limit_cycle_Hemmerich_2,quantum_LC_6,TWA_1,TWA_2,A_M_Rey_TWA}. We demonstrate the signature of bistability between SSR and ASR phases within TWA, where the dynamics is attracted to either of the SR phases depending on the initial conditions. 
%
Furthermore, the self-trapping phenomenon of photons is also verified from the TWA simulation \cite{SM}.
Interestingly, the self-trapped limit cycle also persists in presence of quantum fluctuations up to a certain time scale, which increases with increasing number of atoms. The onset of chaos can also be unveiled from the broad frequency spectrum of individual stochastic trajectories.

Moreover, we confirm these intriguing phenomena from the full quantum dynamics, within the stochastic wavefunction approach \cite{Q_traj_1,Q_traj_2} for a small number of atoms. In the bistable regime of SSR and ASR phases, the quantum trajectories converge to either of the classical steady states (as shown in Fig.\ref{fig2}(b)). Even for small number of atoms, we observe that a fraction of quantum trajectories are attracted towards the self-trapped state, confirming its signature, which we discuss in \cite{SM}.

%

{\it Discussion:} 
The anisotropic Dicke dimer with photon loss displays diverse dynamical phenomena, such as the multistability of nonequilibrium phases, limit cycles, onset of chaos, and quasi-periodic motion. 
Notably, we have identified steady states as well as a new type of limit cycle leading to the self-trapping phenomenon, offering the possibility of observing a unique photonic time crystal.
The saturation value of the decorrelator can serve as an alternate measure for chaoticity. The impact of quantum fluctuations on the dynamical phases deserves further investigation, which may induce a dynamical transition between the coexisting phases. The presence of a chaotic attractor may facilitate chaos-assisted tunnelling between the self-trapped states, opening avenues for exploring the quantum signature of dissipative chaos. 

In conclusion, this atom-photon dimer system offers a platform to study a wide range of intriguing dynamical phenomena, including multistability, photon self-trapping, and dissipative chaos, relevant to experimental setups in cavity and circuit QED with potential applications in quantum technology.

\begin{acknowledgments}
{\it Acknowledgments:} We thank Sudip Sinha for comments and fruitful discussions. D.M. acknowledges support from Prime Minister Research Fellowship (PMRF).
\end{acknowledgments}


\begin{thebibliography}{99}
	
\bibitem{rev_noneq_1}
A. Polkovnikov, K. Sengupta, A. Silva, and M. Vengalattore, Rev. Mod. Phys. {\bf 83}, 863 (2011).	

\bibitem{rev_noneq_2}
J. Eisert, M. Friesdorf, and C. Gogolin, Nat. Phys. {\bf 11}, 124 (2015).

\bibitem{Moore_noneq_3}
R. Vasseur and J. E. Moore, J. Stat. Mech. (2016) 064010.

\bibitem{Huse_noneq_4}
R. Nandkishore and  D. A. Huse,  Annu. Rev. Condens. Matter Phys. {\bf 6}, 15–38 (2015).

\bibitem{rev1}
I. Bloch, J. Dalibard, and W. Zwerger, Rev. Mod. Phys. {\bf 80}, 885 (2008).

\bibitem{rev2}
I. Bloch, J. Dalibard and S. Nascimb\'{e}ne, Nature Phys {\bf 8}, 267–276 (2012).


\bibitem{rev4}
C. Gross and I. Bloch, Science {\bf 357}, 995 (2017).

\bibitem{Serge_Haroche}
J. M. Raimond, M. Brune, and S. Haroche, Rev. Mod. Phys. {\bf 73}, 565 (2001).

\bibitem{Carusotto_Review}
I. Carusotto and C. Ciuti, Rev. Mod. Phys. {\bf 85}, 299 (2013).

\bibitem{Helmut_Ritsch}
F. Mivehvar, F. Piazza, T. Donner, H. Ritsch, Adv. Phys. {\bf 70}, 1–153 (2021).

\bibitem{Esslinger_rev}
H. Ritsch, P. Domokos, F. Brennecke, and T. Esslinger, Rev. Mod. Phys. {\bf 85}, 553 (2013).

\bibitem{Zollar_rev}
M. M\"{u}ller, S. Diehl, G. Pupillo, and P. Zoller, Adv. At. Mol. Opt. Phys. {\bf 61}, 1 (2012).

\bibitem{Steven_Girvin_1}
A. Blais, A. L. Grimsmo, S. M. Girvin, and A. Wallraff, Rev. Mod. Phys. {\bf 93}, 025005 (2021).	

\bibitem{Houck}
A. Houck, H. T\"{u}reci and J. Koch, Nature Phys {\bf 8}, 292–299 (2012).




%

	




\bibitem{Blatter}
S. Schmidt and G. Blatter, Phys. Rev. Lett. {\bf 103}, 086403 (2009).

\bibitem{Le_Hur}
J. Koch and K. Le Hur, Phys. Rev. A {\bf 80}, 023811 (2009).

\bibitem{knap1}
M. Knap, E. Arrigoni, and W. v.d. Linden, Phys. Rev. B {\bf 82}, 045126 (2010).

\bibitem{jcref7}
L. Guo, S. Greschner, S. Zhu, and W. Zhang, Phys. Rev. A {\bf 100}, 033614 (2019).

\bibitem{fazio1}
D. Rossini and R. Fazio, Phys. Rev. Lett. {\bf  99}, 186401 (2007).

\bibitem{Review_ergodicity_1}
L. D’Alessio, Y. Kafri, A. Polkovnikov, and M. Rigol, Adv. Phys. {\bf 65}, 239 (2016).


	


%

\bibitem{Daley_dissipation}
F. Damanet, E. Mascarenhas, D. Pekker, and A. J. Daley, Phys. Rev. Lett. {\bf 123}, 180402 (2019).

\bibitem{Mueller_dissipation}
P.M. Harrington, E.J. Mueller, and K.W. Murch, Nat Rev Phys {\bf 4}, 660–671 (2022).

\bibitem{Dicke_5}
R. Lin, R. Rosa-Medina, F. Ferri, F. Finger, K. Kroeger, T. Donner, T. Esslinger, and R. Chitra, Phys. Rev. Lett. {\bf 128}, 153601 (2022).


\bibitem{Hendric}
H. Weimer, A. Kshetrimayum, and R. Or\'{u}s, Rev. Mod. Phys. 93, 015008 (2021).

\bibitem{Keeling_Dicke}
F. Damanet, A. J. Daley, and J. Keeling, Phys. Rev. A {\bf 99}, 033845 (2019).

\bibitem{Hemmarich}
J. Klinder, H. Keßler, M. Wolke, L. Mathey, and A. Hemmerich, Proc. Natl. Acad. Sci. U.S.A. {\bf 112}, 3290 (2015).

\bibitem{Dis_tran4}
K. C. Stitely, A. Giraldo, B. Krauskopf, and S. Parkins, Phys. Rev. Research {\bf 2}, 033131 (2020).

\bibitem{Dis_tran5}
K. C. Stitely, A. Giraldo, B Krauskopf, and S. Parkins, Phys. Rev. Research {\bf 4}, 023101 (2022).

\bibitem{Dis_tran6}
K. C. Stitely, S. J. Masson, A. Giraldo, B. Krauskopf, and S. Parkins, Phys. Rev. A {\bf 102}, 063702 (2020).



\bibitem{Dis_tran7}
S. Ray, A. Vardi, and D. Cohen, Phys. Rev. Lett. {\bf 128}, 130604 (2022).



\bibitem{Dicke_6}
K. C. Stitely, F. Finger, R. Rosa-Medina, F. Ferri, T. Donner, T. Esslinger, S. Parkins, and B. Krauskopf, Phys. Rev. Lett. {\bf 131}, 143604 (2023).

\bibitem{Dis_tran9}
J. Li, R. Fazio, and S. Chesi, New J. Phys. {\bf 24}, 083039 (2022).


\bibitem{Dis_tran10}
W. Kopylov, M. Radonji\'{c}, T. Brandes, A. Bala\v{z}, and A. Pelster, Phys. Rev. A {\bf 92}, 063832 (2015).

\bibitem{Dis_tran11}
F. Carollo and I. Lesanovsky, Phys. Rev. Lett. {\bf 126}, 230601 (2021).

%
%
%
%
%
%
%


\bibitem{Dis_tran1}
H. J. Carmichael, Phys. Rev. X {\bf 5}, 031028 (2015).

\bibitem{Chitra}
M. Soriente, T. Donner, R. Chitra, and O. Zilberberg, Phys. Rev. Lett. {\bf 120}, 183603 (2018).


\bibitem{Dis_tran8}
C. J. Zhu, L. L. Ping, Y. P. Yang, and G. S. Agarwal, Phys. Rev. Lett. {\bf 124}, 073602 (2020).

\bibitem{Cuiti_dist_trans_1}
A. Le Boit\'{e}, G. Orso, and C. Ciuti, Phys. Rev. Lett. {\bf 110}, 233601 (2013).

\bibitem{Zollar_dist_trans}
S. Diehl, A. Tomadin, A. Micheli, R. Fazio, and P. Zoller, Phys. Rev. Lett. {\bf 105}, 015702 (2010).

\bibitem{dissipative_transition_CQED}
M. Fitzpatrick, N. M. Sundaresan, A. C. Y. Li, J. Koch and A. A. Houck, Phys. Rev. X {\bf 7}, 011016 (2017).

\bibitem{Cuiti_dist_trans_2}
W. Casteels, R. Fazio, and C. Ciuti, Phys. Rev. A {\bf 95}, 012128 (2017).

\bibitem{Milies_dist_trans}
M. H. Kalthoff, D. M. Kennes, A. J. Millis, and M. A. Sentef, Phys. Rev. Research {\bf 4}, 023115 (2022).


\bibitem{supersolid_1}
J. L\'{e}onard, A. Morales, P. Zupancic, T. Esslinger, and T. Donner, Nature (London) {\bf 543}, 87 (2017).

\bibitem{supersolid_2}
J. L\'{e}onard, A. Morales, P. Zupancic, T. Donner, and T. Esslinger, Science {\bf 358}, 1415 (2017).

\bibitem{A_M_Rey_cavity_nature}
J.A. Muniz, D. Barberena, R.J. Lewis-Swan, D. J. Young, J. R. K. Cline, A. M. Rey, Nature {\bf 580}, 602–607 (2020).

\bibitem{A_M_Rey_cavity_exp}
D. J. Young, A. Chu, E. Y. Song, D. Barberena, D. Wellnitz, Z. Niu, V. M. Sch\"{a}fer, R. J. Lewis-Swan, A. M. Rey and  J. K. Thompson,  Nature {\bf 625}, 679–684 (2024).

\bibitem{Dissipative_Rydberg}
F. Letscher, O. Thomas, T. Niederpr\"{u}m, M. Fleischhauer, and H. Ott, Phys. Rev. X {\bf 7}, 021020 (2017).

\bibitem{B_L_Lev}
R. M. Kroeze, Y. Guo, V. D. Vaidya, J. Keeling, and B. L. Lev, Phys. Rev. Lett. {\bf 121}, 163601 (2018).

\bibitem{ADM_1}
Z. Zhiqiang, C. H. Lee, R. Kumar, K. J. Arnold, S. J.
Masson, A. S. Parkins, and M. D. Barrett, Optica {\bf 4}, 424 (2017).

\bibitem{Dicke_exp_1}
M. P. Baden, K. J. Arnold, A. L. Grimsmo, S. Parkins, and M. D. Barrett,
Phys. Rev. Lett. {\bf 113}, 020408 (2014).
%
\bibitem{Dicke_exp_2}
Z. Zhang, C. H. Lee, R. Kumar, K. J. Arnold, S. J. Masson, A. L. Grimsmo, A. S. Parkins, and M. D. Barrett, Phys. Rev. A {\bf 97}, 043858 (2018).
%
\bibitem{Dicke_exp_3}
F. Ferri, R. Rosa-Medina, F. Finger, N. Dogra, M. Soriente, O. Zilberberg, T. Donner, and T. Esslinger, Phys. Rev. X {\bf 11}, 041046 (2021).
%
\bibitem{Stojan_Rebic}
S. Rebi\'{c}, J. Twamley, and G. J. Milburn, Phys. Rev. Lett. {\bf 103}, 150503 (2009).


\bibitem{LC_Disssipative_Rydberg}
D. Ding, Z. Bai, Z. Liu, B. Shi, G. Guo, W. Li, and C. S. Adams, Sci. Adv.{\bf 10},eadl5893 (2024).

\bibitem{continous_time_crystal}
P. Kongkhambut, J. Skulte, L. Mathey, J. G. Cosme, A. Hemmerich, H. Keßler, Science {\bf 377}, 670–673 (2022).

%
\bibitem{Limit_cycle_Hemmerich_2}
J. Skulte, P. Kongkhambut, H. Keßler, A. Hemmerich, L. Mathey, and J. G. Cosme,
Phys. Rev. A {\bf 109}, 063317 (2024).


\bibitem{quantum_LC_6}
H. Keßler, J. G. Cosme, M. Hemmerling, L. Mathey, and A. Hemmerich, Phys. Rev. A {\bf 99}, 053605 (2019).

\bibitem{Quantum_vanderpol}
L. B. Arosh, M. C. Cross, and R. Lifshitz, Phys. Rev. Research {\bf 3}, 013130 (2021).



\bibitem{quantum_LC_4}
Berislav Bu\v{c}a, Cameron Booker, Dieter Jaksch, SciPost Phys. {\bf 12}, 097 (2022). 

\bibitem{quantum_LC_5}
K. Seibold, R. Rota, and V. Savona, Phys. Rev. A {\bf 101}, 033839 (2020).

\bibitem{Dicke_correlator}
L. da Silva Souza, L. F. dos Prazeres, and F. Iemini, Phys. Rev. Lett. {\bf 130}, 180401 (2023).



\bibitem{Haake_chaos_1}
R. Grobe, F. Haake, and Hans-J\"{u}rgen Sommers, Phys. Rev. Lett. {\bf 61}, 1899 (1988).

\bibitem{Shepelyansky}
G. G. Carlo, G. Benenti, and D. L. Shepelyansky, Phys. Rev. Lett. {\bf 95}, 164101 (2005).

\bibitem{Casati_chaos_1}
G. G. Carlo, G. Benenti, G. Casati, and D. L. Shepelyansky, Phys. Rev. Lett. {\bf 94}, 164101 (2005).

\bibitem{Zyczkowsky}
S. Denisov, T. Laptyeva, W. Tarnowski, D. Chru\'{s}ci\'{n}ski, and K. \.{Z}yczkowski, Phys. Rev. Lett. {\bf 123}, 140403 (2019).

\bibitem{Prosen_3}
L. S\'{a}, P. Ribeiro, and T. Prosen, Phys. Rev. X {\bf 10}, 021019 (2020).

\bibitem{BHT_chaos}
D. Dahan, G. Arwas, and E. Grosfeld, npj Quantum Inf {\bf 8}, 14 (2022).

\bibitem{Deb_TCH}
D. Mondal, K. Sengupta, and S. Sinha, Phys. Rev. A {\bf 110}, 042207 (2024).

\bibitem{Minganti}
F. Ferrari, L. Gravina, D. Eeltink, P. Scarlino, V. Savona, F. Minganti, arXiv.2305.15479 (2023).

\bibitem{Russomanno}
G. Passarelli, P. Lucignano, D. Rossini, A. Russomanno, arXiv:2406.16585 (2024).




 




\bibitem{QPT_Chaos_Brandes}
C. Emary and T. Brandes, Phys. Rev. Lett. {\bf 90}, 044101 (2003); 

\bibitem{Sudip_review}
S. Sinha, S. Ray and S. Sinha, J. Phys.: Condens. Matter {\bf 36} 163001 (2024).

\bibitem{Graham_EPL}
T. Dittrich, R. Graham, Europhys. Lett. {\bf 7}(4): 287 (1988).

\bibitem{Kolovsky}
A. R. Kolovsky, Phys. Rev. E {\bf 106}, 014209 (2022).



\bibitem{ADM_2}
P. Nataf, A. Baksic, and C. Ciuti, Phys. Rev. A {\bf 86}, 013832 (2012); A. Baksic and C. Ciuti, Phys. Rev. Lett. {\bf 112}, 173601 (2014).

\bibitem{QPT_0}
R. H. Dicke, Physical Review. {\bf 93} (1): 99–110 (1954).

\bibitem{Haake_2012}
A. Altland and F. Haake, Phys. Rev. Lett. {\bf 108}, 073601 (2012); New J. Phys.
{\bf 14}, 073011 (2012).

\bibitem{Lea_2023}
D. Villase\~{n}or, S. Pilatowsky-Cameo, M. A. Bastarrachea-Magnani, S. Lerma-Hern\'andez, L. F. Santos, and J. G. Hirsch, Entropy {\bf 25}, 8 (2022).

\bibitem{A_M_Rey_Dicke_1}
R.J. Lewis-Swan, A. Safavi-Naini, J. J. Bollinger and A. M. Rey, Nat Commun {\bf 10}, 1581 (2019). 



\bibitem{Scar_Dicke}
S. Pilatowsky-Cameo, D. Villase\~{n}or, M. A. Bastarrachea-Magnani, S. Lerma-Hern\'{a}ndez, L. F. Santos and J. G. Hirsch, Nat Commun {\bf 12}, 852 (2021).

\bibitem{Scar_sudip}
S. Sinha and S. Sinha, Phys. Rev. Lett. {\bf 125}, 134101 (2020).

\bibitem{Dicke_1}
F. Mivehvar, Phys. Rev. Lett. {\bf 132}, 073602 (2024).

\bibitem{Dicke_2}
E. I. R. Chiacchio, A. Nunnenkamp, and M. Brunelli, Phys. Rev. Lett. {\bf 131}, 113602 (2023).

\bibitem{JCH_Plenio}
M. Hartmann, F. G. S. L. Brand\~{a}o and M. B. Plenio, Nature Phys {\bf 2}, 849–855 (2006).

\bibitem{npj_hopping}
M. Kounalakis, C. Dickel, A. Bruno, N. Langford, and G. Steele, npj Quantum Inf. {\bf 4}, 38 (2018).



\bibitem{Obarthalar_1}
M. Albiez, R. Gati, J. F\"{o}lling, S. Hunsmann, M. Cristiani, and M. K. Oberthaler, Phys. Rev. Lett. {\bf 95}, 010402 (2005); Th. Anker, M. Albiez, R. Gati, S. Hunsmann, B. Eiermann, A. Trombettoni, and M. K. Oberthaler, Phys. Rev. Lett. {\bf 94}, 020403 (2005).

\bibitem{Oberthalar_2}
T. Zibold, E. Nicklas, C. Gross, and M. K. Oberthaler, Phys. Rev. Lett. {\bf 105}, 204101 (2010).

\bibitem{AC_DC}
S. Levy, E. Lahoud,  I. Shomroni, J. Steinhauer, Nature {\bf 449}, 579–583 (2007).


\bibitem{Self_trapping1}
M. Abbarchi, A. Amo, V. G. Sala, D. D. Solnyshkov, H.Flayac, L. Ferrier, I. Sagnes, E. Galopin, A. Lema\^{i}tre, G. Malpuech, and J. Bloch, Nat. Phys. {\bf 9}, 275 (2013).

\bibitem{JC_dimer_expt}
J. Raftery, D. Sadri, S. Schmidt, H. E. T\"{u}reci, and A. A. Houck, Phys. Rev. X {\bf 4}, 031043 (2014).

\bibitem{Shenoy1}
A. Smerzi, S. Fantoni, S. Giovanazzi, and S. R. Shenoy, Phys. Rev. Lett. {\bf 79}, 4950 (1997);  S. Raghavan, A. Smerzi, S. Fantoni, and S. R. Shenoy, Phys. Rev. A {\bf 59}, 620 (1999).
%
%
\bibitem{Shenoy2}
K. Sakmann, A. I. Streltsov, O. E. Alon, and L. S. Cederbaum, Phys. Rev. Lett. {\bf 103}, 220601 (2009).

\bibitem{Sebastian_ST}
S. W\"{u}ster, Beata J. D\c{a}browska-W\"{u}ster, and M. J. Davis, Phys. Rev. Lett. {\bf 109}, 080401 (2012).


\bibitem{JCD_Houck}
S. Schmidt, D. Gerace, A. A. Houck, G. Blatter, and H. E. T\"{u}reci, Phys. Rev. B {\bf 82}, 100507(R) (2010).

\bibitem{Vivek_JCD}
G. Vivek, Debabrata Mondal, and S. Sinha, Phys. Rev. E {\bf 108}, 054116 (2023).

\bibitem{Manas_TCD}
T. Ray, M. Kulkarni, Phys. Rev. A {\bf 110}, 032220 (2024).
%

\bibitem{Shenoy_damping}
I. Marino, S. Raghavan, S, Fantoni, S. R. Shenoy and A. Smerzi, Phys. Rev. A {\bf 60} 487 (1999).



\bibitem{Breuer}
H. P. Breuer and F. Petruccione, {\it The Theory of Open Quantum Systems} (Oxford University Press, Oxford, 2007).

\bibitem{Drossel}
B. Fernengel and B. Drossel, J. Phys. A: Math. Theor. {\bf 53} 385701 (2020).

\bibitem{SM}
See the Supplementary material for the details of the LG potentials, multistability, onset of chaos and effect of photon hopping, decay rate, detuning as well as the quantum results in the present model.

\bibitem{foontnote}
Both the symmetric and anti-symmetric classes satisfy the condition $s_{zL} = s_{zR}$ as a consequence of the exchange and parity symmetries.

\bibitem{Lorentz}
Lorenz, E. N. (1963). Deterministic Nonperiodic Flow. Journal of the Atmospheric Sciences, {\bf 20} (2), 130-141.


\bibitem{Strogatz}
S. H. Strogatz, {\it Nonlinear Dynamics and Chaos} (Westview Press, Boulder, CO, 2007).

\bibitem{Lichtenberg}
A. J. Lichtenberg and M. A. Lieberman, {\it Regular and chaotic dynamics}, (Springer-Verlag,1992).

\bibitem{Lyapunov}
J. Ch\'{a}vez-Carlos, M. A. Bastarrachea-Magnani, S. Lerma-Hern\'{a}ndez, and J. G. Hirsch, Phys. Rev. E {\bf 94}, 022209 (2016).

\bibitem{Decorr1}
A. Das, S. Chakrabarty, A. Dhar, A. Kundu, D. A. Huse, R. Moessner, S. Sankar Ray, and S. Bhattacharjee, Phys. Rev. Lett. {\bf 121}, 024101 (2018).

\bibitem{Decorr2}
V. Khemani, D. A. Huse, and A. Nahum,  Phys. Rev. B {\bf 98}, 144304 (2018).

\bibitem{Decorr3}
T. Bilitewski, S. Bhattacharjee, and R. Moessner, Phys. Rev. B {\bf 103}, 174302 (2021).

\bibitem{Decorr4}
S. Ruidas and S. Banerjee, SciPost Phys. {\bf 11}, 087 (2021).

\bibitem{Decorr5}
A. K. Chatterjee, A. Kundu, and M. Kulkarni, Phys. Rev. E {\bf 102}, 052103 (2020).

\bibitem{Decorr6}
A. Deger, S. Roy, and A. Lazarides,  Phys. Rev. Lett. {\bf 129}, 160601 (2022).

\bibitem{Decorr7}
A. Schuckert, M. Knap, SciPost Phys. {\bf 7}, 022 (2019).

\bibitem{TWA_1}
P. B. Blakie, A. S. Bradley, M. J. Davis, R. J. Ballagh, and C. W. Gardiner, Adv. Phys. {\bf 57}, 363 (2008).
%
\bibitem{TWA_2}
A. Polkovnikov, Ann. Phys. {\bf 325}, 1790 (2010).
%
\bibitem{A_M_Rey_TWA}
J. Huber, A. Maria Rey, and P. Rabl, Phys. Rev. A {\bf 105}, 013716 (2022).

\bibitem{Q_traj_1}
H. Carmichael, {\it An Open Systems Approach to Quantum Optics}. Springer-Verlag  (1993).
%
\bibitem{Q_traj_2}
K. Mølmer, Y. Castin and J. Dalibard, J. Opt. Soc. Am. B {\bf 10}, 524 (1993).


\end{thebibliography}
\end{document}


\begin{center}
	{\bf SUPPLEMENTARY MATERIAL:\\
		Self-trapping phenomenon, multistability and chaos in open anisotropic Dicke dimer}
\end{center}

\section{STEADY STATES AND THEIR STABILITY ANALYSIS}
We obtain the fixed points $\mathbf{X}^*$ representing the steady states from the equations of motion (given in Eq.(3) of the main text) by setting $\dot{\mathbf{X}}=0$.
To investigate the stability of the different phases of the open anisotropic Dicke dimer (ADD), we perform a linear stability analysis of the equations of motion  around the fixed points representing the steady states. We consider fluctuation in the dynamical variables around the FPs, namely, $\mathbf{X}(t) = \mathbf{X}^* +\delta \mathbf{X}(t)$, where $\delta\mathbf{X}(t) = \delta \mathbf{X} e^{i\omega_s t}$. In the dissipative system, $\omega_s$ can be complex and the stability of a steady state is ensured by the condition ${\rm Re}(\omega_s)<0$ \cite{Strogatz}. By following this procedure, we have obtained the stability regimes of the various phases in the parameter space, as given in the phase diagram (see the main text for details). 

\section{Effective Landau-Ginzburg potential}

In this section, we discuss the dissipative transitions between the normal to superradiant phase (SR) and the formation of self-trapped states in the framework of effective Landau-Ginzburg (LG) theory, which can also capture the nature of their transitions.
From the EOM (see Eq.(3) of the main text) the steady states with finite photon number (with $x_i,p_i\ne 0$) correspond to the condition of vanishing determinant,
\begin{center}
	$\begin{vmatrix}
		\Omega_{L+}& \kappa \omega_0 & -J \omega_0 & 0 \\
		-\kappa \omega_0 & \Omega_{L-} & 0 & -J \omega_0 \\
		-J \omega_0 & 0 & \Omega_{R+} & \kappa\omega_0 \\
		0 & -J \omega_0 & -\kappa \omega_0 & \Omega_{R-} 
	\end{vmatrix}$
	= 0,
\end{center}
where, $\Omega_{i\pm}=\omega \omega_0+(\lambda_- \pm \lambda_+)^2s_{zi}$.
In terms of the variables $s_{z\pm}=(s_{z\rm L}\pm s_{z\rm R})/2$, the above determinent reduces to the algebric equation,
\begin{eqnarray}
	A_1 s_{z+}^4+A_2 s_{z-}^4+A_3 s_{z+}^3+A_4 s_{z+}^2+A_5 s_{z-}^2+A_6 s_{z+}+A_7 s_{z+}^2s_{z-}^2+A_8 s_{z-}^2s_{z+}+ A_9 = 0,
	\label{General_equations}
\end{eqnarray}
where the coefficients are given by,
\begin{subequations}
	\begin{eqnarray}
		A_1 &=& A_2 = (\lambda_-^2-\lambda_+^2)^4;\quad A_3 = 4\omega\omega_0(\lambda_-^2-\lambda_+^2)^2(\lambda_-^2+\lambda_+^2)\\
		A_4 &=& 2\omega_0^2(\omega^2+\kappa^2)(\lambda_-^2-\lambda_+^2)^2+4\omega^2\omega_0^2(\lambda_-^2+\lambda_+^2)^2-2J^2\omega_0^2(\lambda_-^4+6\lambda_-^2\lambda_+^2+\lambda_+^4)\\
		A_5 &=& 2\omega_0^2(\omega^2+\kappa^2)(\lambda_-^2-\lambda_+^2)^2-4\omega^2\omega_0^2(\lambda_-^2+\lambda_+^2)^2+2J^2\omega_0^2(\lambda_-^4+6\lambda_-^2\lambda_+^2+\lambda_+^4)\\
		A_6 &=& 4\omega\omega_0^3(\lambda_-^2+\lambda_+^2)(\kappa^2+\omega^2-J^2);\quad A_7 = -2(\lambda_-^2-\lambda_+^2)^4; \quad A_8 = -4\omega\omega_0(\lambda_-^2-\lambda_+^2)^2(\lambda_-^2+\lambda_+^2)\\
		A_9 &=& \omega_0^4[J^4+(\kappa^2+\omega^2)^2+2J^2(\kappa^2-\omega^2)].
	\end{eqnarray}	
\label{Coefficients}
\end{subequations}
Although from the dynamical point of view, the transitions correspond to the bifurcation of the fixed points, below we show how they can also be understood from an effective LG framework even in the presence of dissipation. The condition given in Eq.\eqref{General_equations} corresponds to the minima of the  LG potential describing the symmetry broken phases.

\subsection{Superradiant transition}
As mentioned in the main text, the superradiant phases SSR and ASR belonging to the symmetric and anti-symmetric classes ($s_{z-}=0$), originate through the bifurcation of the normal phase NP$_{\downarrow\downarrow}$. 
\begin{figure}[h]
	\begin{center}
		\includegraphics[width = 0.7\linewidth]{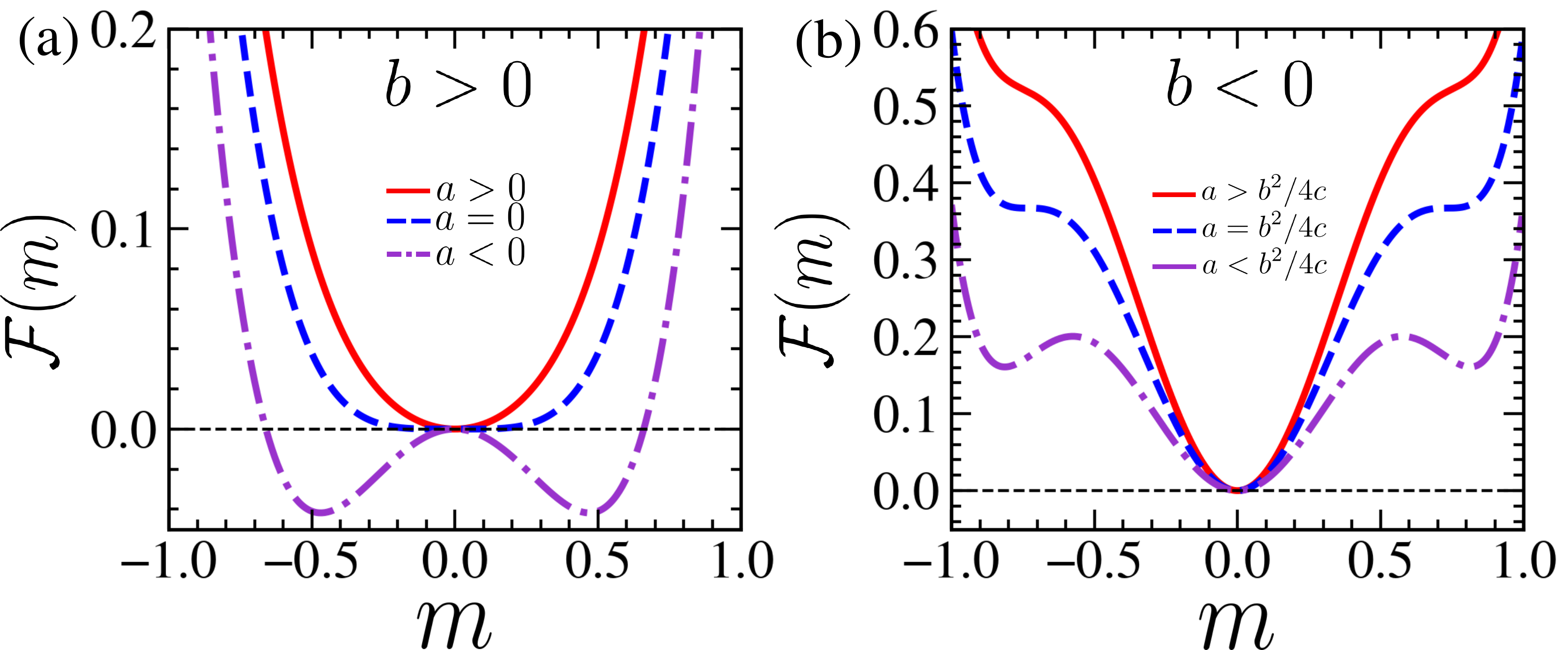}
		\caption{{\it Effective Landau-Ginzburg potential describing the superradiant transition:} Nature of the potential for (a) the second order and (b) first order like transition between normal phase NP$_{\downarrow\downarrow}$ and superradiant phase in the different parameter regimes. The interaction strengths $\lambda_{\pm}$ are measured in the units of hopping amplitude $J$. We set $\hbar,k_B = 1$ and $\omega=2.7,\omega_0=0.9,\kappa = 0.9$ for all figures.} 
		\label{fig1}    
	\end{center}
\end{figure}
The normal phase is characterized by zero photon number and $s_{zi}=-1$ and after the transition, superradiant phase arises with non-vanishing photon number and  $|s_{zi}|<1$, due to which the spin polarization can serve as an effective order parameter to capture the superradiant transition. 
Such non-equilibrium transition and its nature can be described by using an effective Landau-Ginzburg potential $\mathcal{F}(m)=\frac{a}{2}m^2+\frac{b}{4}m^4+\frac{c}{6}m^6$ with the coefficients,
\begin{subequations}
	\begin{eqnarray}
		a &=& (\lambda_{-}^2-\lambda_{+}^2)^2-2\tilde{\omega}\omega_0(\lambda_{-}^2+\lambda_{+}^2)+\omega_0^2(\tilde{\omega}^2+\kappa^2),\qquad\\
		b &=& 2\tilde{\omega}\omega_0(\lambda_{-}^2+\lambda_{+}^2)-2(\lambda_{-}^2-\lambda_{+}^2)^2\\
		c &=& (\lambda_{-}^2-\lambda_{+}^2)^2,
	\end{eqnarray}
\end{subequations}	
where the order parameter can be defined as $m^2=1+s_{z+}$ and $\tilde{\omega}=\omega\mp J$ corresponding to the symmetric and anti-symmetric classes respectively.   The minima of this LG potential describing the superradiant phases satisfies the Eq.\eqref{General_equations}.

In the parameter regime characterized by $b>0$, the potential $\mathcal{F}(m)$ changes its nature from a single well to a double well at the critical point $a=0$ (see Fig.\ref{fig1}(a)), indicating a continuous transition. On the other hand, for $b<0$, the nature of the potential changes and a new local minima arises, which finally gives rise to the first order like transition at $a=b^2/(4c)$, due to which the superradiant phase appears with the jump in the photon number. From the point of view of the nonlinear dynamics, this scenario corresponds to a saddle-node bifurcation as discussed in the main text and the change in the nature of the transition at $a=b=0$ serves as a tricritical point in the thermodynamic sense.
Note that, this formalism only captures the transition and its nature, however, the stability of the phases can be analyzed from the propagation of small amplitude fluctuation.

\subsection{Dissipative transition of the self-trapped state}
The open ADD model exhibits self-trapping phenomenon due to the breaking of exchange symmetry between the two cavities. 
\begin{figure}[h]
	\begin{center}
		\includegraphics[width = 0.7\linewidth]{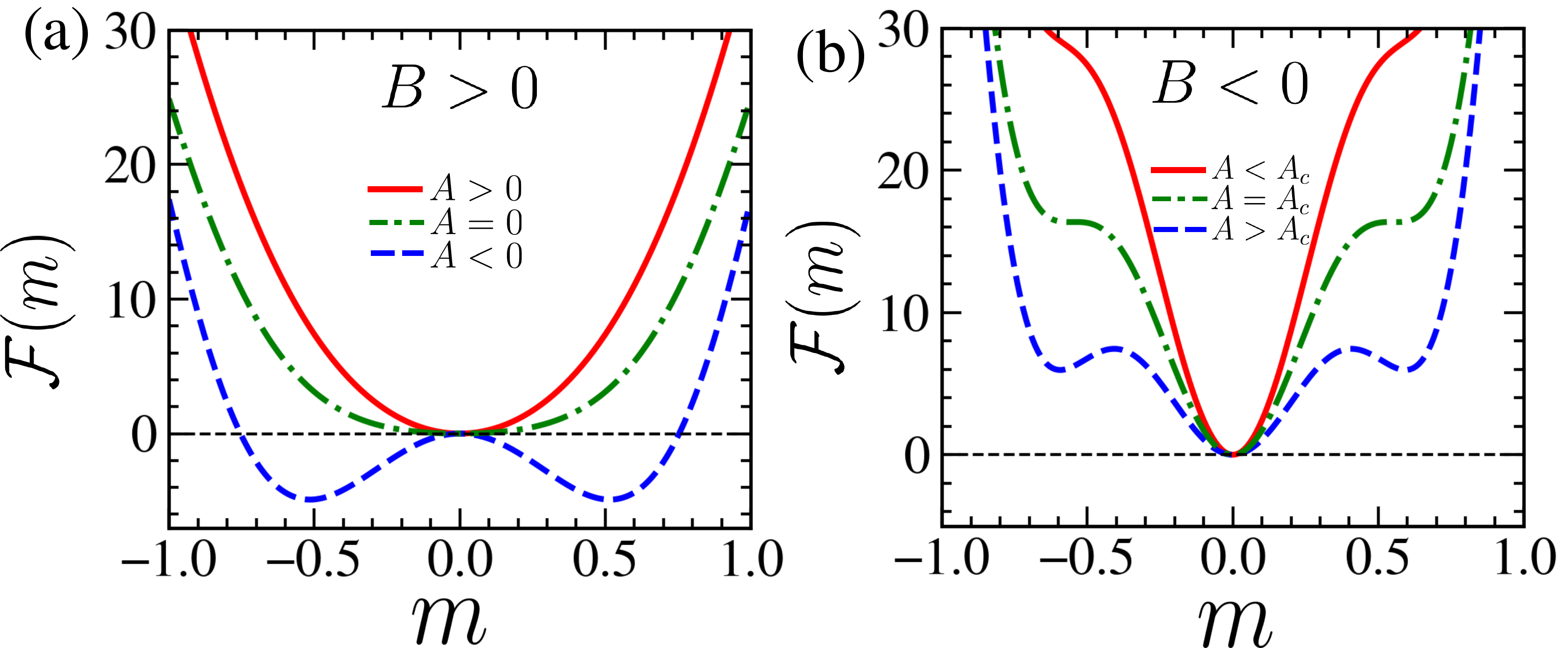}
		\caption{{\it Effective Landau-Ginzburg potential describing ST transition:} Nature of the potential across (a) the continous and (b) first order like transition to self-trapped state for different parameter regimes.} 
		\label{fig2}    
	\end{center}
\end{figure}
We numerically find that the self-trapped (ST) states mainly arise from unstable normal phases NP$_{\uparrow\downarrow}$ or NP$_{\downarrow\uparrow}$ with $s_{zL}=-s_{zR}=\pm 1$, as discussed in the main text.
The nature of the nonequilibrium transition that gives rise to this ST state can be understood from the following effective LG potential,
\begin{equation}
	\mathcal{F}(m)=A\frac{m^2}{2}+B\frac{m^4}{4}+C\frac{m^6}{6}-D\frac{m^8}{2}+D\frac{m^{10}}{10},
\end{equation}
where one of the order parameters can be defined as $m^2 = 1-s_{z-}$ which takes a non-vanishing value at the ST state.
The coefficients can be written in terms of the components $A_j$ (see Eq.\eqref{Coefficients}) and are given by, 
\begin{subequations}
	\begin{eqnarray}
		A &=& A_1 s_{z+}^4+A_3 s_{z+}^3+(A_4+A_7)s_{z+}^2+(A_8+A_6) s_{z+}+A_2+A_5+A_9\\
		B &=& -(4A_2+2A_5+2A_7s_{z+}^2+2A_8s_{z+})\\
		C &=& 6A_2+A_5+A_7s_{z+}^2+A_8s_{z+}; \\
		D &=& A_2,
	\end{eqnarray}
\end{subequations}
which depend on the coupling constants and $s_{z+}$ obtained numerically. The ST states described by the minima of this potential satisfies the Eq.\eqref{General_equations}. Similar to the superradiant transition, for $B >0$, the potential changes its nature from single well to double well across $A = 0$, exhibiting second order transition.  Whereas, a first order transition occurs for $B<0$ at critical point $A=A_c$ describing a saddle-node bifurcation. It is important to mention that the stable ST phase arises only through the saddle-node bifurcation as depicted in the phase diagram given in Fig.3(a) of the main text.

\section{Coexistence of multiple stable steady states}
\begin{figure}[h]
	\begin{center}
		\includegraphics[width = 0.95\linewidth]{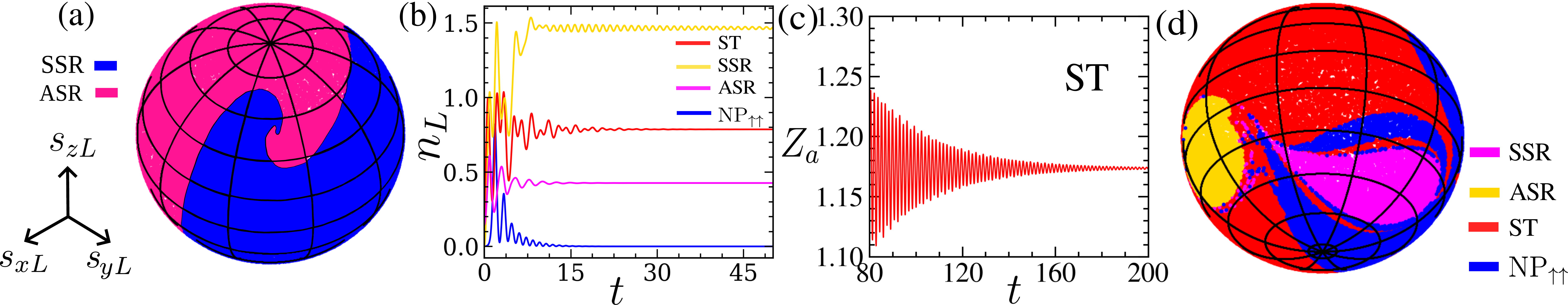}
		\caption{{\it Multistability of different steady states:} (a) Basins of attraction of the different superradiant phases SSR and ASR in the bistable regime at $\lambda_-=2.2$, $\lambda_+=1.2$ (see the main text for details).
			(b) Dynamics of the photon number $n_L$ of the left cavity for different initial conditions, leading to different phases. For ST state, we show only the branch with higher photon number. 
			(c) The dynamics of the atomic population imbalance $Z_a = |s_{z\rm L}|-|s_{z\rm R}|$ for ST state. 
			(d) The basin of attraction over the Bloch sphere of one of the cavities, corresponding to the different states in the multi-stability regime. To obtain the basin of attraction on the spin space of the left cavity, we have fixed the other dynamical variables to certain random values. We set the parameters: $\lambda_-=0.8$, $\lambda_+=2.9$.} 
		\label{fig3}    
	\end{center}
\end{figure}

\textbf{Bistability of superradiant phases:} As discussed in the main text, there is a regime in the phase diagram (see Fig.2(a) of the main text) where both types of superradiant phases SSR and ASR coexist resulting in bistability. As shown in Fig.\ref{fig3}(a), these superradiant phases have disjoint basins of attraction projected on the Bloch sphere of one of the cavities, which guides us to choose the initial condition appropriately.

\textbf{Multistability in the regime of self-trapped state:} It is evident from the phase diagram given in Fig.3(a) of the main text, that there exists a region in the parameter space where four distinct steady states can coexist, indicating multistability. Similar to the bistability of the superradiant phases SSR, ASR discussed in the main text, in this region additionally the normal phase NP$_{\uparrow\uparrow}$ as well as the self-trapped (ST) state coexist. 
In such a region of multistability, a complex dynamical behavior is expected and depending on the initial condition, the trajectories are attracted to the different fixed points, which is shown over the Bloch sphere given in Fig.3(d) of the main text.  
The corresponding dynamics of the photon number of the left cavity $n_L$ is shown in Fig.\ref{fig3}(b), which clearly distinguishes the different fixed points. 
The main distinguishing feature of the self-trapped state is the nonzero photon population imbalance between the two cavities $|Z_p| > 0$, as shown in Fig.3(b1) of the main text. Similarly, the atomic population imbalance $Z_a=|s_{z\rm L}|-|s_{z\rm R}|$ also yields nonzero saturation value for ST state, which is evident from its time evolution, as shown in Fig.\ref{fig3}(c).  
To probe the different phases corresponding to steady states in the regime of multistability, the initial condition should be chosen appropriately.
As depicted in Fig.\ref{fig3}(d), the different fixed points have non-overlapping basins of attraction projected on the Bloch sphere of one of the cavities.

\section{Onset of chaos and its detection using Decorrelator}
\begin{figure}[H]
  \begin{center}
  \includegraphics[width = \linewidth]{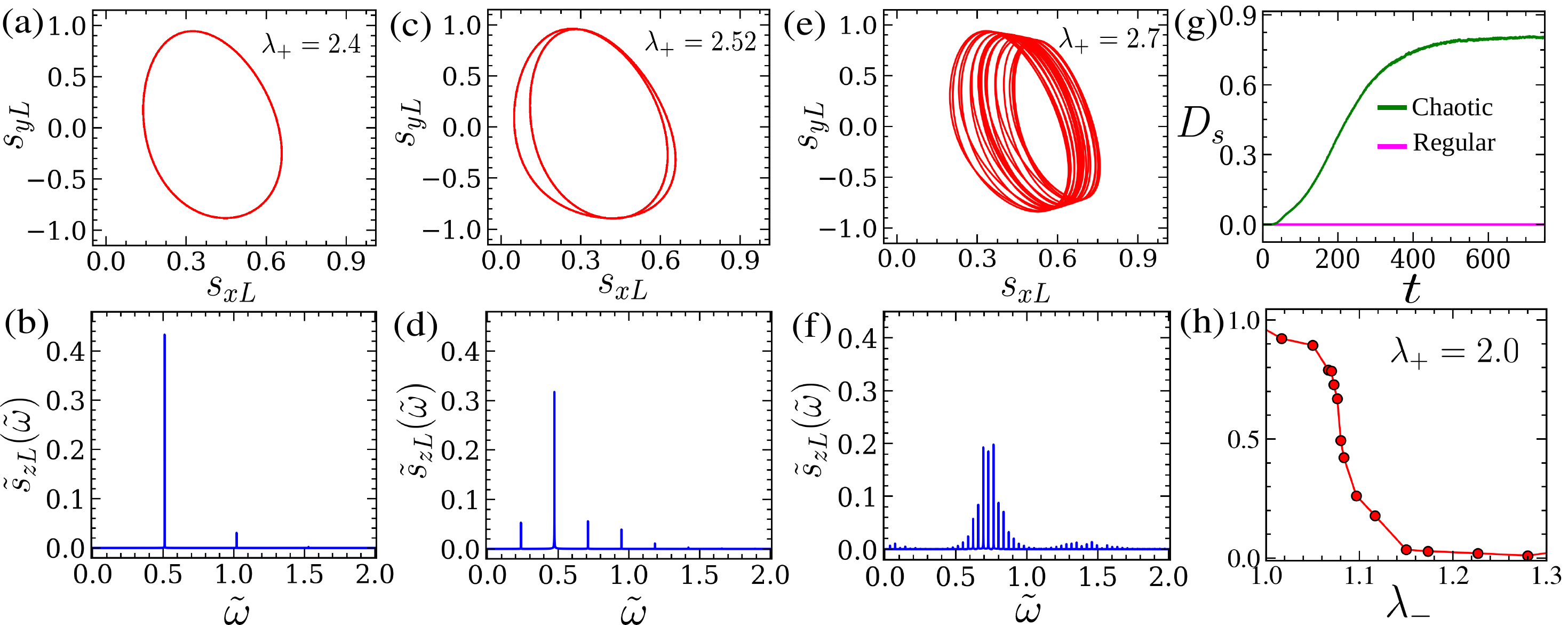}
  \caption{ {\it Instability of the limit cycle and the quasi-periodic motion:} The period doubling bifurcation of limit cycle SLC leading to quasi-periodic state shown from (a,c,e) the trajectory on $s_x-s_y$ plane and (b,d,f) corresponding Fourier spectrum. (g) The dynamics of Decorrelator with $\lambda_-=1.0$ in regular ($\lambda_+=1.2$) versus chaotic regime ($\lambda_+=1.8$). (h) Saturation value $D_s$ of decorrelator as a function of $\lambda_-$ for $\lambda_+=2.0$, showing crossover from chaotic to quasi-periodic behavior.} 
  \label{fig4}    
  \end{center}
\end{figure}

The emergence of chaotic dynamics can be observed in the ADD model even in the presence of dissipation arising due to the photon loss. Above the critical line of the superradiant transition, there is a narrow region in the phase diagram where the region of regular dynamics shrinks significantly due to the instability of the fixed points. However, in a tiny region of the phase diagram, the ST state coexists with chaos (see Fig.5 of the main text). 
The chaotic dynamics gradually transforms to quasi-periodic motion (shown in Fig.\ref{fig4}(e)) with the peaked distribution of incommensurate frequencies (see Fig.\ref{fig4}(f)) and low Lyapunov exponent. Such quasiperiodic dynamics (indicated by light blue colour in Fig.4(a) of the main text) spreads over a large region of the phase diagram.
The symmetric limit cycle SLC becomes unstable and the dynamics is attracted to the quasi-periodic phases, for which one sharp frequency corresponding to the limit cycle transforms into a peaked distribution of frequencies characterizing the quasi-periodic dynamics (shown in Fig.\ref{fig4}(a-f)).
Such crossover from unstable limit cycle to quasi-periodic motion resembles a period-doubling phenomenon, as depicted in Fig.\ref{fig4}(a-f). On the contrary, the onset of chaos from the limit cycle phase occurs more sharply. Note that, such quasi-periodic region is absent in the single cavity anisotropic Dicke model.

Chaotic dynamics can be detected using the Lyapunov exponent (LE), which measures the sensitivity of the initial condition. Typically, for a chaotic trajectory, a small initial perturbation $\delta \mathbf{X}(t=0)$ at the phase space point $\mathbf{X}$ grows exponentially in time $||\delta\mathbf{X}(t)||=e^{\Lambda t} \,||\delta\mathbf{X}(0)||$, which yields the Lyapunov exponent \cite{Strogatz} as,
\begin{eqnarray}
	\Lambda&=&\lim_{t \to \infty}\frac{1}{t}\ln \left(\frac{||\delta\mathbf{X}(t)||}{||\delta\mathbf{X}(0)||}\right).
	\label{LE}
\end{eqnarray}
We compute the mean Lyapunov exponent $\bar{\Lambda}$ averaged over an ensemble of initial phase space points to quantify the  degree of chaoticity, as shown in Fig.4(a) of the main text. Alternatively, the chaos can be measured from the dynamics of the Decorrelator, a classical analogue of the out-of-time-order correlator \cite{Decorr1,Decorr2}. In this work, we have studied the Decorrelator dynamics of the spin in one of the cavities. It is defined as $D(i,t)=1-\langle{\bf s}_i^a\cdot {\bf s}_i^b\rangle$,
where $a$ and $b$ denote two copies of the initial spin configurations (of both the cavities), with the $i$th spin of the configuration $b$ is chosen to be slightly perturbed from $a$. Here, the averaging is performed over an ensemble of initial phase space points. In the chaotic regime, the Decorrelator grows exponentially and saturates to a nonzero value $D_s$, whereas in the regular region, it vanishes, as depicted in Fig.\ref{fig4}(g). Furthermore, in the chaotic regime, the saturation value $D_s$ of the Decorrelator increases, as shown in Fig.4(b) of the main text, which serves as an alternate measure to detect chaos. The comparison between the saturation value $D_s$ and average Lyapunov exponent $\bar{\Lambda}$ is shown in Fig.4(b) of the main text. Furthermore, the variation of the saturation value $D_s$ of the decorrelator from the chaotic to quasi-periodic region is shown in Fig.\ref{fig4}(h).

\section{Complete phase diagram illustrating all the dynamical phases}
\begin{figure}[H]
	\begin{center}
		\includegraphics[width = 0.7\linewidth]{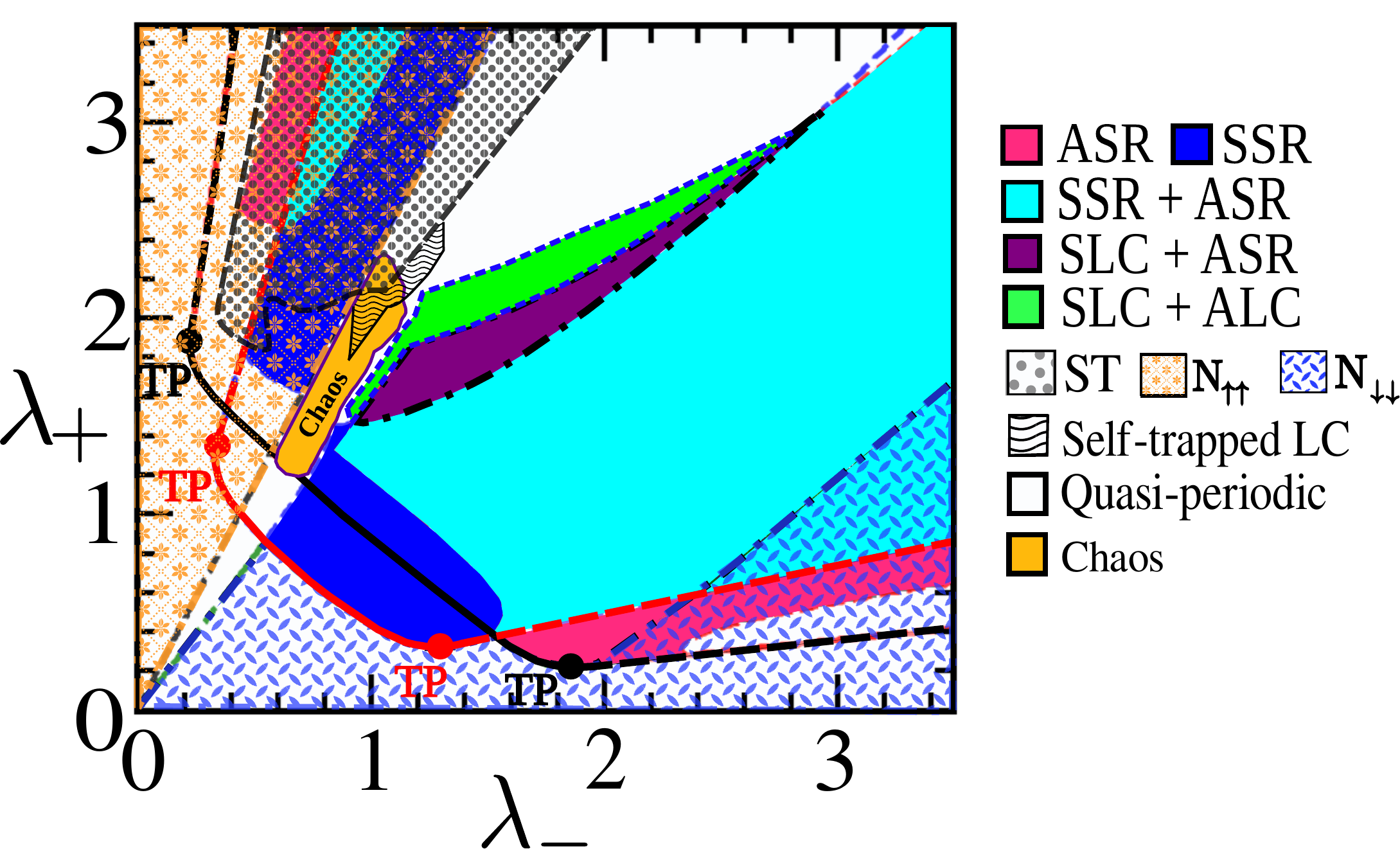}
		\caption{Dynamical phase diagram in $\lambda_{-}-\lambda_{+}$ plane, depicting the stability regimes of all phases discussed in the present work. Parameter chosen: $\omega=2.7,\omega_0=0.9,J=1.0, \kappa=0.9$.} 
		\label{fig5}    
	\end{center}
\end{figure}

In this section, we present the complete phase diagram of the anisotropic Dicke dimer in Fig.\ref{fig5}, including normal phases, superradiant phases, self-trapped state, limit cycles, self-trapped limit cycle, quasi-periodic motion and the onset of chaos. We particularly focus on the stability regimes of stationary states to demonstrate the bistability and multistability phenomena of different superradiant phases and self-trapped state. Moreover, the phase diagram also highlights the coexistence of nonstationary phases such as different limit cycles (SLC, ALC). 
The detailed description of these phases and the corresponding dynamics of the physical quantities in their respective regimes are discussed in the main text. 

\section{Dependence of dynamical phases on various tunable parameters}
In this section, we investigate the influence of different tunable parameters, particularly the photon decay rate $\kappa$ and coupling strength $J$ between the cavities on various stationary phases. For this purpose, we mainly focus on the superradiant phases, especially their bistable regime, and the stable self-trapped state, exploring how these phenomena depend on the variation of $J$ and $\kappa$.

\begin{figure}[t]
	\begin{center}
		\includegraphics[width = 0.7\linewidth]{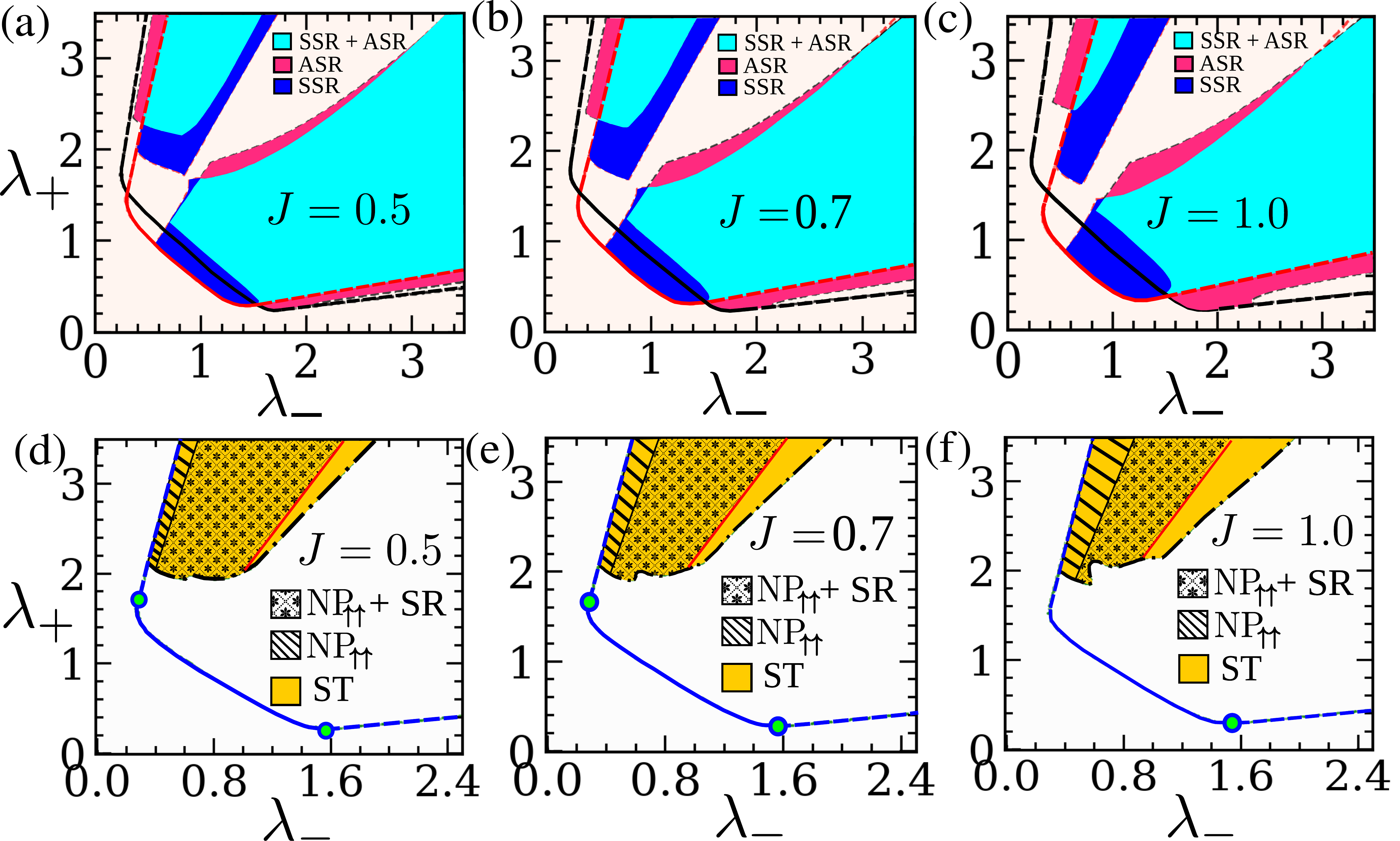}
		\caption{Variation of the stability regimes of the (a,b,c) superradiant phases SSR (blue), ASR (pink) and their bistability (cyan) as well as (d,e,f) ST state (yellow) and its multi-stability  with superradiant phases and normal phase NP$_{\uparrow\uparrow}$ (shaded regions)
		in $\lambda_{-}-\lambda_{+}$ plane with photon hopping strength $J$. Parameter chosen: $\omega=2.7,\omega_0=0.9, \kappa=0.9.$} 
		\label{fig6}    
	\end{center}
\end{figure}

\begin{figure}[H]
	\begin{center}
		\includegraphics[width = 0.7\linewidth]{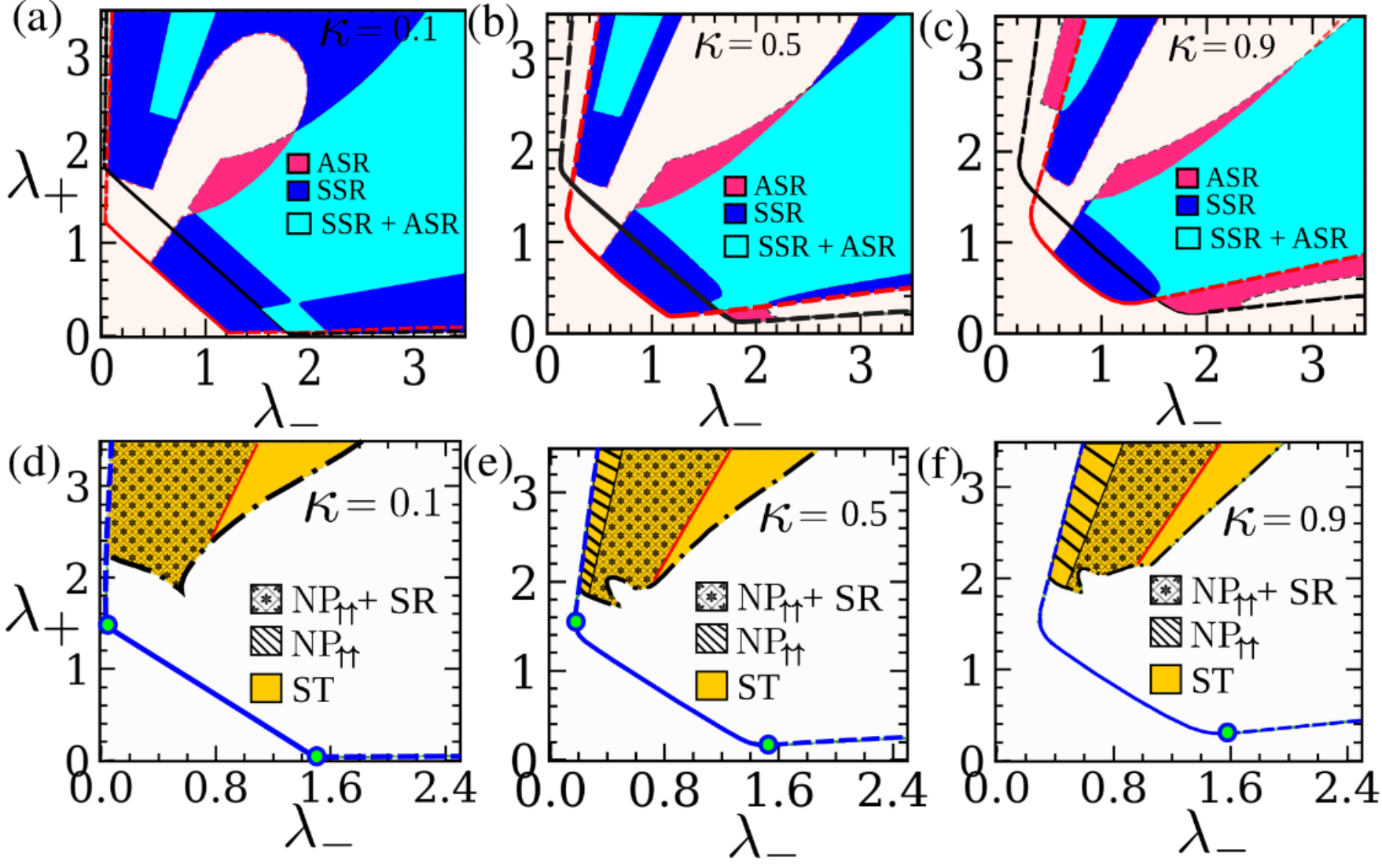}
		\caption{Variation of the stability regimes of the (a,b,c) superradiant phases SSR (blue), ASR (pink) and their bistability (cyan) as well as (d,e,f) ST state (yellow) and its multi-stability  with superradiant phases and normal phase NP$_{\uparrow\uparrow}$ (shaded regions)
		in $\lambda_{-}-\lambda_{+}$ plane with photon decay rate $\kappa$. Parameter chosen: $\omega=2.7,\omega_0=0.9, J=1.0.$} 
		\label{fig7}    
	\end{center}
\end{figure}

The phase diagrams of the superradiant phases, illustrating their bistability for different values of the coupling strength $J$ and photon decay rate $\kappa$, are shown in Figs.\ref{fig6}(a, b, c) and \ref{fig7}(a, b, c), respectively [similar to Fig.2(a) of the main text]. Furthermore, the stability regime of the self-trapped state and its dependence on $J$ and $\kappa$ are depicted in the phase diagrams shown in Figs.\ref{fig6}(d, e, f) and \ref{fig7}(d, e, f), respectively [similar to Fig.3(a) of the main text].
It is evident from these figures that the qualitative behavior of the phase diagrams remains the same as shown in Fig. 2(a) and Fig. 3(a) of the main text, even with variations in the tunable parameters $J$ and $\kappa$. Although the superradiant phases remain stable over a large region of the parameter space, their stability regions in the top-left part of the phase diagrams $(\lambda_+>\lambda_-)$ shrink as $J$ and $\kappa$ increase.

As reflected in this analysis, the bistability region of the superradiant phases and the stability regime of the self-trapped state remain largely unaffected by variations in the photon decay rate $\kappa$ and photon hopping strength $J$, ensuring the robustness of these phenomena.


\subsection{Effect of detuning on the self-trapped state and onset of chaos}
\begin{figure}[H]
	\begin{center}
		\includegraphics[width = 0.8\linewidth]{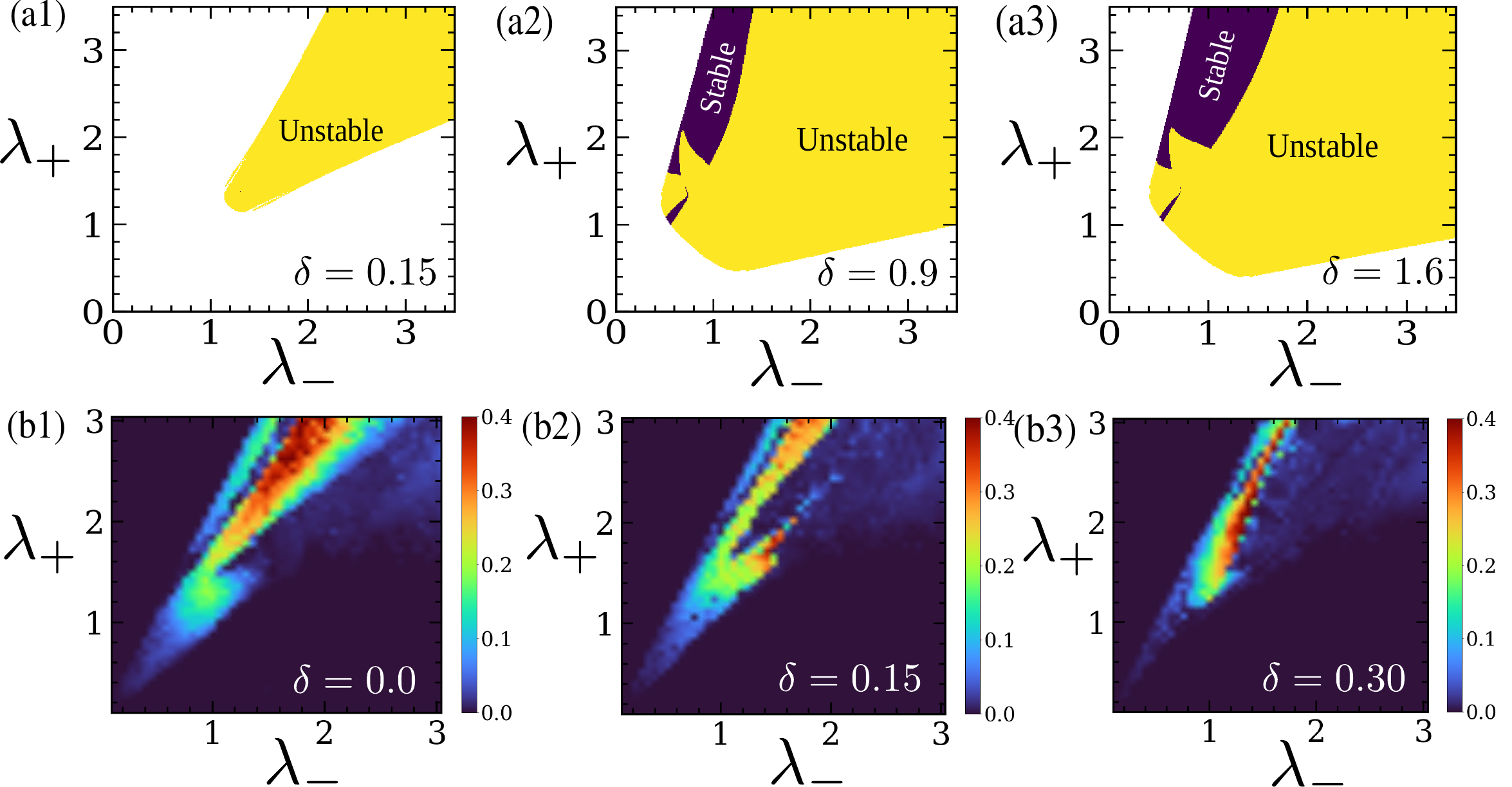}
		\caption{ {\it Effect of detuning in open ADD:} (a1-a3) Growth of the stability region of self-trapped state and (b1-b3) suppression of the chaotic region from the color scale plot of LE in the parameter space with increasing detuning $\delta$. } 
		\label{fig8}    
	\end{center}
\end{figure}

In ADD, detuning $\delta = \omega-\omega_0$ between the frequencies of atoms and photons plays an important role in controlling the dynamical behavior. We find that the self-trapping phenomenon ceases to exist at a small value of the detuning as the ST state becomes unstable (see Fig.\ref{fig8}(a1)). 
It is evident from Fig.\ref{fig8}(a1-a3) that the detuning favours the self-trapping phenomenon as its stability region increases with detuning.  

In contrast, the overall chaotic region is suppressed with increasing detuning (see Fig.\ref{fig8}(b1-b3)), however, it exists in a small region in the parameter space.  Therefore, the coexistence of a self-trapped state and chaos can only be observed by carefully choosing the value of detuning.

\section{Quantum dynamics}
To this end, we incorporate quantum fluctuations using two different methods, which confirm the validity of the results obtained in the classical limit.

\subsection{Truncated Wigner approximation}
In atom-cavity systems, quantum fluctuations are suppressed as the number of atoms increases since effective Planck constant scales as $\hbar/N$.
For a sufficiently large number of atoms $N\approx10^4$, we perform truncated Wigner approximation (TWA) \cite{TWA_1,TWA_2,TWA_3} to describe the dynamics of the open ADD model.
%
%
To mimic quantum fluctuations, we sample an ensemble of initial conditions from a Gaussian distribution around a phase-space point $\{\alpha_i,s_{zi},\phi_i\}$ and evolve them using the following dynamical equations,
\begin{subequations}
	\begin{eqnarray}
         \dot{\alpha_i} &=& -(\kappa+\imath\omega) \alpha_i-\frac{\imath}{\scalebox{0.8}{$\sqrt{2}$}}(\lambda_-s_{i}^-+\lambda_+s_{i}^+)+\imath J\alpha_{\bar{i}}+\sqrt{\frac{\kappa}{2S}}\,\xi \\
	\dot{s}_{i}^+ &=& \imath\omega_0 s_{i}^+-\imath\scalebox{0.8}{$\sqrt{2}$} s_{zi}(\lambda_-\alpha_i^*+\lambda_+\alpha_i)\\
	\dot{s}_{zi} &=& -\frac{\imath}{\scalebox{0.8}{$\sqrt{2}$}}[ \lambda_-(\alpha_i s_{i}^+-\alpha_i^*s_{i}^-)+\lambda_+(\alpha^*_i s_{i}^+-\alpha_is_{i}^-)]
	\end{eqnarray}
	\label{EOM}
\end{subequations}
where $\xi(t)$ describes stochastic quantum noise arising due to the photon loss, which satisfies $\langle \xi^*(t)\xi(t')\rangle=\delta(t-t')$ \cite{TWA_3}. Each initial states are evolved separately up to a sufficiently long time and finally, we obtain the ensemble average of any observable as $\langle O\rangle_{\scriptscriptstyle\text{TWA}} = \sum_j O_j(t)/{\mathcal{N}}$, where $\mathcal{N}$ is the number of trajectories with index $j$. 

\begin{figure}[H]
	\begin{center}
		\includegraphics[width = \linewidth]{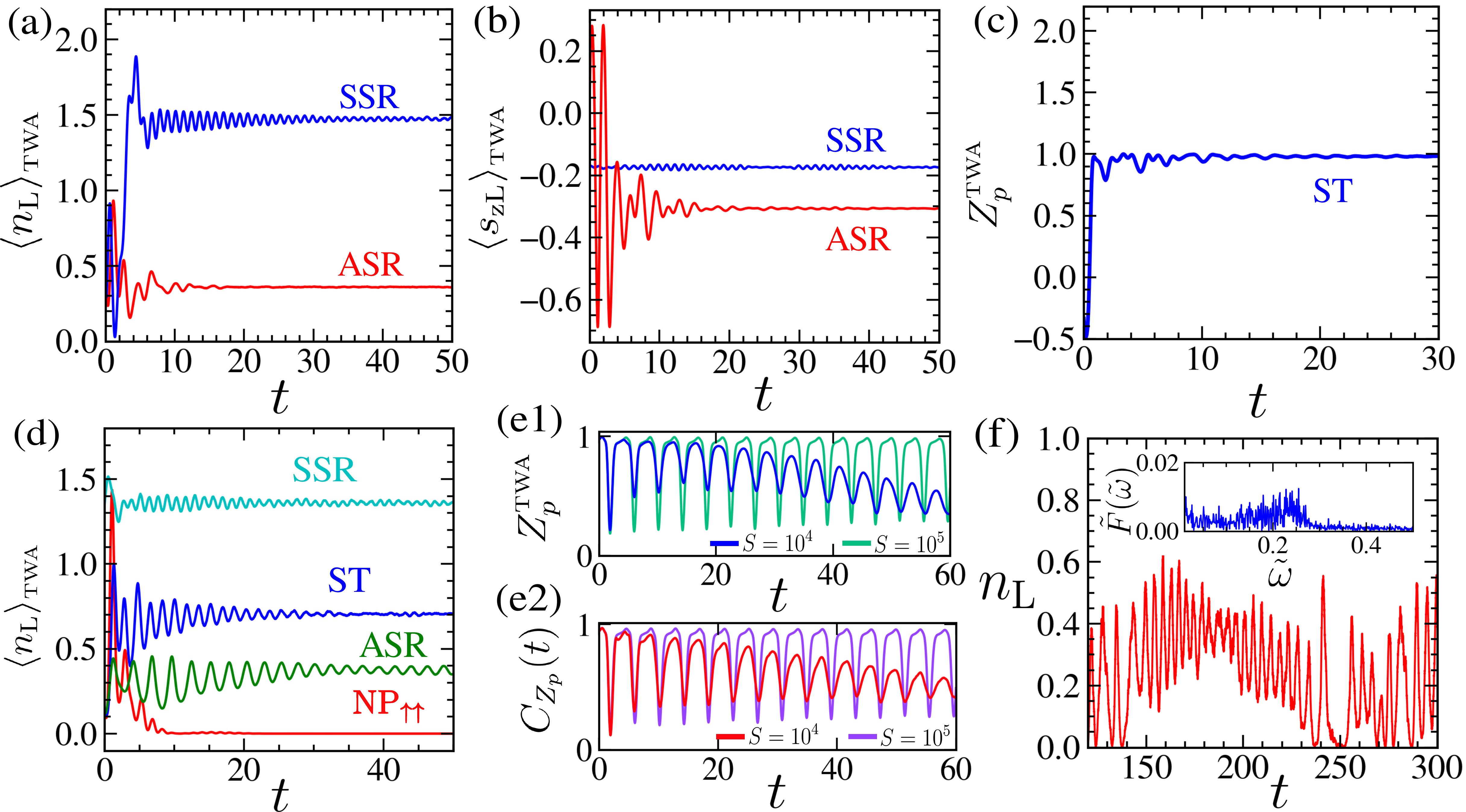}
		\caption{ {\it Semiclassical dynamics within truncated Wigner approximation:}  dynamics of (a) photon number $\langle n_{\rm L}\rangle_{\scriptscriptstyle\text{TWA}}$ and (b) spin polarization $\langle s_{z{\rm L}}\rangle_{\scriptscriptstyle\text{TWA}}$ in the bistable regime of superradiant phases SSR and ASR at $\lambda_{-}=2.2,\lambda_{+}=1.2$ (see phase diagram given in Fig.2(a) of main text). The evolution of (c) photon population imbalance $Z_p^{\scriptscriptstyle\text{TWA}}$ corresponding to the self-trapped state and (d) multistability from the dynamics of photon number $\langle n_{\rm L}\rangle_{\scriptscriptstyle\text{TWA}}$ for $\lambda_-=0.8$, $\lambda_+=2.7$. Time evolution of (e1) photon imbalance $Z_p^{\scriptscriptstyle\text{TWA}}$ and (e2) correlation function $C_{Z_p}(t)$ for the self-trapped limit cycle at $\lambda_-=1.15$, $\lambda_+=2.12$ for spin magnitude $S=10^4,10^5$. (f) Chaotic dynamics of single stochastic trajectory in terms of photon number $n_{\rm L}$ for $\lambda_-=0.7$, $\lambda_+=1.4$. The inset shows corresponding Fourier spectrum. We consider magnitude of the spins $S=10^4$ for a,b,c,d,f.  } 
		\label{fig9}    
	\end{center}
\end{figure}

First, we study the semiclassical dynamics in the bistable regime of superradiant phases SSR and ASR (shown in the phase diagram, given in Fig.2(a) of the main text) using the TWA simulation.
%
%
As evident from Fig.\ref{fig9}(a,b), the dynamics of physical quantities such as photon number $\langle n\rangle_{\scriptscriptstyle\text{TWA}}$ and spin polarization $\langle s_{z}\rangle_{\scriptscriptstyle\text{TWA}}$ converge to the steady state values of SSR and ASR phases depending on the initial conditions guided by respective basins of attraction. 

We also investigate the self-trapping of photons in the corresponding regime of the phase diagram (see Fig.3(a) of the main text) within TWA. To characterize the ST state, we examine the semiclassical dynamics of average photon population imbalance $Z_p^{\scriptscriptstyle\text{TWA}}=\langle \frac{n_{\rm L}-n_{\rm R}}{n_{\rm L}+n_{\rm R}}\rangle$ by properly selecting the initial condition, which also converges to the corresponding fixed point obtained classically (see Fig.\ref{fig9}(c)). Interestingly in this regime of multistability, the normal phase NP$_{\uparrow\uparrow}$, as well as both the superradiant phases coexist with ST state, which is verified from the TWA simulation (see Fig.\ref{fig9}(d)).
%
%
Furthermore, the self-trapped limit cycle also persists in the presence of quantum fluctuations up to a certain time scale, which
increases with increasing number of atoms, as seen from the dynamics of photon imbalance $Z_p^{\scriptscriptstyle\text{TWA}}$ in Fig.\ref{fig9}(e1). The temporal coherence of the limit cycle can also be quantified from the correlation function $C_{Z_p}(t)=\langle Z_p(t)Z_p(0)\rangle$ of the photon number imbalance $Z_p$, depicted in Fig.\ref{fig9}(e2). 

The onset of chaos can also be unveiled from the fluctuations of individual stochastic trajectories. 
%
The spreading of the frequency spectrum of such a typical trajectory signals the chaotic behavior \cite{Lichtenberg}, as shown in Fig.\ref{fig9}(f). 

\subsection{Quantum trajectories}
Next, we investigate the full quantum dynamics for a small number of atoms $N=10$ in open ADD by solving the master equation (given in Eq.(2) in the main text), using the stochastic wavefunction approach \cite{Q_traj_1,Q_traj_2}. To reduce the numerical complexity, we focus on the evolution of a sample of quantum trajectories, starting from a product coherent state $\ket{\Psi_c}=\prod_i\ket{\alpha_i}\otimes\ket{\theta_i,\phi_i}$ of photon and spin degree of freedom in each cavity, representing the appropriate phase-space point $\{\alpha_i,s_{zi}=\cos\theta_i,\phi_i\}$. 

\begin{figure}[H]
	\begin{center}
		\includegraphics[width = 0.7\linewidth]{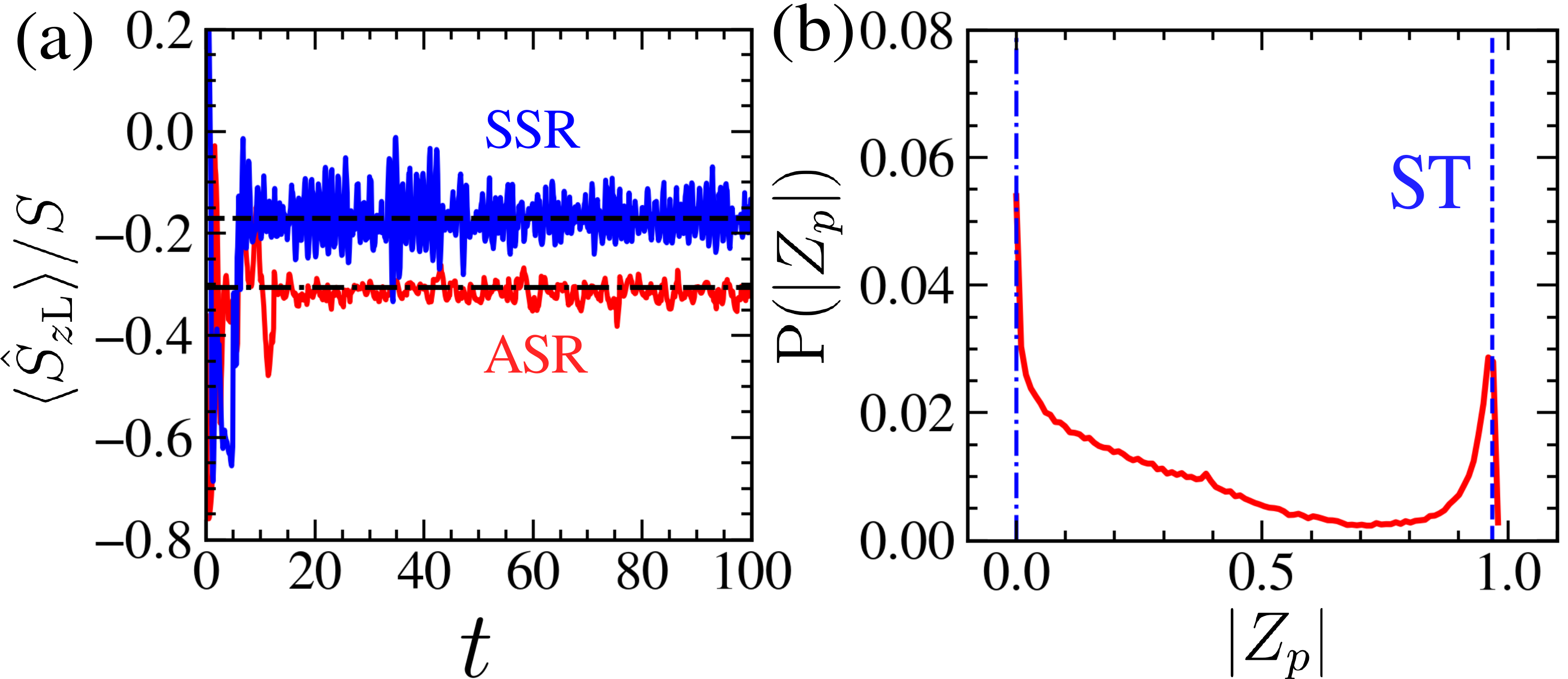}
		\caption{{\it Quantum dynamics within stochastic wavefunction approach:} dynamics of spin polarization $\langle \hat{S}_{z{\rm L}}\rangle/S$ along single quantum trajectories attracted toward the SSR or ASR phases in the bistable regime at $\lambda_{-}=2.2,\lambda_{+}=1.2$. Dashed and dashed dotted lines represent the classical steady states. (b) Distribution of the photon imbalance $|Z_p|$ over the quantum trajectories revealing signature of normal phase NP$_{\uparrow\uparrow}$ and self-trapped state ST in the multistable regime for $\lambda_-=0.8$, $\lambda_+=2.7$. We consider the spin magnitude: $S=5$.} 
		\label{fig10}    
	\end{center}
\end{figure}

We explore the dynamics in the bistable regime of two types of superradiant phases SSR and ASR from the individual quantum trajectories. 
%
The physical quantities like photon number $\langle\hat{n}_i\rangle$ and spin polarization $\langle\hat{S}_{zi}\rangle$ for different quantum trajectories converge to either of the SSR or ASR phase with small fluctuations around them, confirming the bistability (as shown in Fig.2(b) of the main text and Fig.\ref{fig10}(a)). For comparison with the classical results, we scale the average quantities obtained quantum mechanically by $S$. 
%

Moreover, we detect the signature of the self-trapping phenomenon of photons in the multistable regime of the phase diagram, starting from the coherent state representing the corresponding self-trapped steady state.
We find that during the stochastic quantum evolution, a fraction of quantum trajectories remain close to self-trapped steady state.
For better understanding, we obtain the distribution of photon imbalance $Z_p$ corresponding to different trajectories, which exhibits 
a peak at a nonzero photonic imbalance $|Z_p|\approx 1$ revealing the signature of ST state (see Fig.\ref{fig10}(b)). 
Since ST state coexists with normal phase NP$_{\uparrow\uparrow}$ and other superradiant phases without any photon imbalance we find another peak in the distribution near $Z_p= 0$.

%
%


